\documentclass[amsmath,12pt,amssymb,preprint,nofootinbib,aps]{revtex4-2}
\usepackage{amsfonts} 
\usepackage{graphicx} 
\usepackage{epsfig}
\usepackage{multirow}
\usepackage{bm}
\usepackage{color}
\usepackage{amssymb}
\usepackage{dsfont}
\usepackage{hyperref}
\usepackage{cleveref}
\usepackage[margin=1in]{geometry}
\usepackage{longtable}
\usepackage{booktabs}
\usepackage{array}

\begin{document}
	\title{Impact of anisotropy on QCD phase structure and transport coefficients of quark matter}
	
	\author{Dhananjay Singh}
	\email{snaks16aug@gmail.com}
	\affiliation{Department of Physics, Dr. B R Ambedkar National Institute of Technology Jalandhar, 
		Jalandhar -- 144008, Punjab, India}
	
	\author{Arvind Kumar}
	\email{kumara@nitj.ac.in}
	\affiliation{Department of Physics, Dr. B R Ambedkar National Institute of Technology Jalandhar, 
		Jalandhar -- 144008, Punjab, India}	
	
	\begin{abstract}
		Employing the Polyakov chiral SU(3) mean field (PCQMF) model, we investigate how momentum-space anisotropy, characteristic of quark-gluon plasma (QGP) in ultrarelativistic heavy-ion collisions (uRHIC), impacts the thermodynamic behavior and transport coefficients of strongly interacting quark matter. The momentum anisotropy is introduced via a small deformation in the momentum distribution, quantified by a spheroidal parameter $\xi$, which deforms the distribution functions and captures anisotropic effects to linear order. The PCQMF model captures key non-perturbative aspects of QCD, like chiral symmetry breaking, deconfinement dynamics through Polyakov loop potential, and is extended here to accommodate momentum-space anisotropy. We compute the modifications induced by momentum-space anisotropy to key thermodynamic observables including pressure $p$, energy density $\epsilon$, entropy density $s$, speed of sound squared $c_s^2$, and specific heat $c_v$, alongside key transport coefficients, such as shear viscosity $\eta$, bulk viscosity $\zeta_b$, and electrical conductivity $\sigma_{el}$. These coefficients are derived using the relativistic Boltzmann equation (RBE) under the relaxation time approximation (RTA). We find that even a weak anisotropy can lead to significant modifications in the thermodynamic response and transport behavior of quark matter. This underscores the importance of including momentum anisotropy for realistic modeling of the QCD medium across all energy regimes, from current studies at RHIC and LHC to future explorations of the high-density frontier at FAIR, NICA, and J-PARC.
		
	\end{abstract}
	\maketitle
	\newpage
	
	\section{\label{intro}Introduction}
	Ultrarelativistic heavy-ion collision (uRHIC) experiments, carried out at the Relativistic Heavy Ion Collider (RHIC) and the Large Hadron Collider (LHC), have been instrumental in creating and studying the QGP, a state of matter where quarks and gluons are deconfined \cite{star,phenix,alice,brahms}. The asymmetric initial geometry in non-central collisions and rapid expansion of the QGP lead to anisotropic pressure gradients, and the momentum distribution of emitted particles becomes anisotropic in the local rest frame \cite{strick14,plumari19}. In particular, the medium expands more along one direction than others, leading to unequal pressures (and particle yields) in different azimuthal directions, and this anisotropy may persist through much of the QGP’s evolution \cite{romat}. This anisotropy is reflected in the momentum distribution of produced particles and quantified using Fourier coefficients $v_n$ of the azimuthal distribution of emitted particles. The second harmonic (elliptic flow $v_2$) and third harmonic (triangular flow $v_3$) have been measured extensively at RHIC and LHC, providing direct evidence of anisotropic collective expansion \cite{alice2}. This indicates that the QGP, behaving nearly as an ideal fluid, retains memory of the initial geometry through much of its evolution. 
	
	Momentum anisotropy is expected to influence the thermodynamic and transport properties of QCD matter, especially at finite temperature and non-zero chemical potentials. While lattice QCD (LQCD) calculations indicate that the transition from QGP to hadrons is a smooth crossover at low baryon density \cite{aoki,bernard,hot,hot2,hot3,borsanyi,borsanyi2}, the notorious sign problem limits LQCD at high density \cite{split}, so one must rely on QCD low-energy effective models to explore this regime. 
	In the literature, these models are widely used to explore the phase structure of dense matter, and they generally predict the existence of a first-order chiral phase transition terminating in a critical end point (CEP) \cite{fuku,ratti,costa,ferreira,schaefer,skokov}. Experimental facilities such as NICA at JINR (Russia) \cite{jinr}, FAIR at GSI (Germany), including its CBM \cite{cbm} and PANDA \cite{panda} experiments, as well as J-PARC (Japan) \cite{jparc}, aim to provide critical insights into the properties of QCD matter at high baryon densities. Effective models provide a controlled framework to include essential QCD features such as spontaneous chiral symmetry breaking and confinement. These include the quark-meson coupling (QMC) model \cite{qmc,qmc2}, the Polyakov-loop extended QMC (PQMC) model \cite{pqm}, the Nambu-Jona-Lasinio (NJL) model \cite{njl,njl2}, the Polyakov NJL (PNJL) model \cite{pnjl,pnjl2}, the linear sigma model (LSM) \cite{lsm}, its Polyakov extended version PLSM \cite{plsm}, the chiral SU(3) quark mean field model (CQMF) \cite{cqmf}, and Polyakov CQMF (PCQMF) model \cite{pcqmf_manisha}. In addition, other approaches such as the functional renormalization group approach \cite{frg,frg2} and Dyson-Schwinger equation approach \cite{dse} have also been proposed to probe the behaviour of QCD at finite chemical potentials. 
	These models have been widely applied to investigate the thermodynamic \cite{schaefer09} and transport properties \cite{abhi,ghosh19} of strongly interacting matter.
	
	Recent model developments account for effects such as strong magnetic and electric fields \cite{fuku10,gatto11,kashiwa,tawfik15,kamal25,ruggieri,tavares,kamal24}, finite volume \cite{bhatt13,zhang20,magdy19,abreu,palhares,zhao19,shaikh}, non-extensivity \cite{zhao20,rath24}, and vorticity \cite{wang19,hernandez} to faithfully capture the complexity of the QCD matter produced in uRHIC. Conventionally, these models have relied on the assumption that quark matter remains perfectly isotropic in momentum space in the absence of magnetic fields. However, realistic QGP is inherently anisotropic due to initial geometric asymmetry and unequal expansion dynamics along the longitudinal (beam) and transverse directions of the medium. Initial spatial anisotropy causes significantly lower pressure along the longitudinal direction, resulting in faster expansion and cooling along the beam direction \cite{romat}. As a result, the transverse momenta of particles dominate significantly over longitudinal momenta, making the parton momentum distributions anisotropic \cite{baier}. Hydrodynamic and AdS/CFT analyses, as well as classical Yang–Mills simulations within
	the color-glass-condensate framework \cite{strick14,heller,schee,mclerran}, all indicate that the momentum anisotropy persists for at least $\tau \lesssim 2$ fm/c, with pressure ratios $P_L/P_T$ well below unity. These results show that a realistic phenomenological analysis of strongly interacting matter must incorporate momentum anisotropy.  
	
	The anisotropic momentum distributions not only modifies the local properties of the QGP but also induces chromo-Weibel instabilities associated with color collective modes, driving rapid redistribution of energy and momentum, thereby shaping the QGP’s early-time evolution and subsequent hydrodynamic behavior \cite{randrup,arnold05,dumitru,romat06}. In practice, momentum anisotropy is typically modelled by a parameter $\xi$, which encodes the degree of deformation along a specified direction, effectively quantifying the extent to which the distribution is stretched or compressed relative to isotropy. This approach, first developed by Romatschke and Strickland \cite{romat}, has since become a central tool in kinetic and hydrodynamic treatments of anisotropic QGP. Numerous studies have explored how momentum anisotropy alters key observables, from photon and dilepton yields \cite{schenke,bhatt16,kasmaei} to parton self-energies \cite{schenke06,kasmaei16}, heavy-quark potentials \cite{nopo,dumitru08,prakash}, quarkonium states \cite{burnier,margotta,lata13}, jet-quenching coefficients \cite{giat}, and fundamental transport coefficients \cite{srivastava,he}, as well as non-equilibrium hydrodynamic behavior \cite{alga}.
	
	Momentum anisotropy can significantly modify the equation of state and the transport properties of the QGP. Transport coefficients, such as shear viscosity $\eta$, bulk viscosity $\zeta_b$, and electrical conductivity $\sigma_{el}$, which govern the QGP's response to gradients and external fields, are known to be particularly sensitive to anisotropic effects. Recent studies have explored these effects on QCD matter using various effective models. For instance, a 2+1 flavor quark-meson (QM) model showed that increasing the anisotropy parameter, $\xi$, suppresses $\eta$ and $\sigma_{el}$ and leads to a nontrivial behavior in $\zeta_b$ \cite{zhang21}. A similar reduction in viscosity and conductivity was confirmed using a quasiparticle relaxation time approximation (RTA) to the Boltzmann equation \cite{lata17}. Within the PNJL model, Ref. \cite{he} found that anisotropy amplified baryon-number fluctuations, enhancing kurtosis, skewness, and deforming isentropic trajectories. Ref. \cite{srivastava} showed that $\sigma_{el}$ increases with temperature but is suppressed by anisotropy in a quasiparticle RTA setup.
	 In a three-flavor PQM analysis impact of finite anisotropy on nature of phase transition has been explored in Ref.  \cite{nisha24}. 
	The above various studies underscore the importance of incorporating momentum-space anisotropy for accurately describing the thermodynamic properties and transport coefficients of QCD matter.
	
	Motivated by this, we aim to examine the influence of momentum anisotropy on the thermodynamic observables and transport coefficients of strongly interacting matter within the Polyakov chiral SU(3) quark mean field (PCQMF) model, across a wide range of temperatures and chemical potentials relevant to heavy-ion collisions. The PCQMF model is an effective framework that incorporates both chiral and deconfinement dynamics by including Polyakov loop variables in a three-flavor chiral mean field model. It contains quark interactions with scalar and vector meson fields, which drive chiral symmetry breaking and its restoration, along with a Polyakov loop potential that effectively captures gluon dynamics for confinement. This approach effectively reproduces QCD thermodynamics across a range of temperatures and chemical potentials \cite{manisha21}. It has been used to study the fluctuations of conserved charges \cite{nisha22}, nonextensive \cite{dj24}, finite-volume effects \cite{dj25}, and properties of nuclear as well as quark stars \cite{manisha22,manisha22q}. The momentum-space anisotropy is incorporated via a dimensionless parameter $\xi$, which quantifies the extent of deformation along a specific direction, thereby capturing deviations from local isotropy in the medium. Employing the anisotropic PCQMF framework, we compute key thermodynamic quantities, including pressure, energy density, entropy density, speed of sound, and specific heat. We then calculate transport coefficients, such as shear viscosity, bulk viscosity, and electrical conductivity, by solving the relativistic Boltzmann equation (RBE) within the RTA under anisotropic conditions. We assume an isotropic relaxation time $\tau$, following standard practice,
	and discuss the implications and limitations of this assumption in the section \ref{transport}. It should be noted that implementing anisotropy affects only the momentum dependence within distribution functions and does not introduce new fundamental interactions. Additionally, we restrict ourselves to a weakly anisotropic medium, i.e., $|\xi|<1$, allowing us to expand the anisotropic distribution function linearly around its isotropic equilibrium form. This enables us to decompose the transport coefficients explicitly into isotropic contributions and anisotropic corrections \cite{srivastava}. 
	
	The remainder of this paper is structured as follows. In Sec. \ref{PCQMF}, we outline the theoretical framework of the PCQMF model. The incorporation of momentum-space anisotropy into the PCQMF model is presented in Sec. \ref{anisotropy}. Anisotropic expressions of the transport coefficients are derived in Sec. \ref{transport}. In Sec. \ref{results}, we examine the influence of momentum anisotropy on the thermodynamic and transport properties of the medium, with comparisons to available lattice QCD results. Finally, Sec. \ref{summary} summarizes our findings and offers physical interpretations of the observed effects.

	\section{\label{method} Methodology }
	In this section, we detail the theoretical framework employed to investigate the impact of momentum anisotropy on the thermodynamic and transport properties of quark matter and its implications on the QCD phase diagram. We begin with an overview of the PCQMF model. Then, we outline the procedure for incorporating momentum anisotropy into the model and derive the anisotropic expressions for shear viscosity, bulk viscosity, and electrical conductivity by employing the RBE within RTA.
	
	\subsection{\label{PCQMF} The PCQMF Model}
	The PCQMF model is an effective QCD-based framework that incorporates key non-perturbative aspects of QCD, specifically, spontaneous breaking of chiral symmetry and quark confinement, within a thermodynamically consistent mean-field approximation  \cite{pcqmf_manisha}. The Lagrangian density is expressed as
	\begin{equation}
		\mathcal{L}_{\text{PCQMF}} = \mathcal{L}_{\text{chiral}} - {\cal U}(\Phi,\bar{\Phi},T).
		\label{lag_pcqmf}
	\end{equation}
	Here, the term $\mathcal{L}_{\text{chiral}}$ contains the interactions between quarks and mesonic fields, along with self-interaction terms that govern spontaneous chiral symmetry breaking and its subsequent restoration at finite temperature and density. The quantity ${\cal U}(\Phi,\bar{\Phi},T)$ denotes the effective Polyakov loop potential, which encodes the deconfinement transition dynamics via the traced Polyakov loop variable $\Phi$ and its conjugate $\bar{\Phi}$. The chiral Lagrangian is decomposed as \cite{wang03}
	\begin{equation}
		\mathcal{L}_{\text{chiral}} = \mathcal{L}_{q0} + \mathcal{L}_{qm} + \mathcal{L}_{M} + \mathcal{L}_{\Delta m},
		\label{lag_chiral}
	\end{equation}
	where $\mathcal{L}_{q0} = \bar{\psi}(i\gamma^\mu \partial_\mu) \psi$ describes the kinetic part of the free quark fields $\psi = (u,d,s)$. The quark-meson interactions are incorporated by the term $\mathcal{L}_{qm}$ which is given by
	\begin{eqnarray}
		{\cal L}_{qm}  =  g_s\left(\bar{\psi}_LM \psi_R+\bar{\psi}_RM^+\psi_L\right)- g_v\left(\bar{\psi}_L\gamma^\mu l_\mu \psi_L+\bar{\psi}_R\gamma^\mu r_\mu \psi_R\right),
		\label{lag_qm} 
	\end{eqnarray}
	where $g_s$ and $g_v$ denote the scalar and vector coupling constants, respectively, determining the strength of the quark interactions with the corresponding mesonic fields. The scalar(pseudoscalar) and vector(axial-vector) meson nonets are defined as $M(M^+) = \frac{1}{\sqrt{2}} \sum_{a=0}^8 (\sigma_a \pm i\pi_a)\lambda_a$ and $l_\mu(r_\mu) = \frac{1}{2\sqrt{2}} \sum_{a=0}^8 (v_\mu^a \pm r_\mu^a)\lambda_a$, respectively, with $\lambda_a$ being the Gell-Mann matrices. The dynamics of the meson sector are described by the term $\mathcal{L}_{M}=\mathcal{L}_{\Sigma \Sigma}+\mathcal{L}_{VV}+\mathcal{L}_{SB}$. The first term $\mathcal{L}_{\Sigma \Sigma}$, details the self-interactions of the scalar mesons and is written as
	\begin{eqnarray}
		{\cal L}_{\Sigma\Sigma} =& -\frac{1}{2} \, k_0\chi^2
		\left(\sigma^2+\zeta^2+\delta^2\right)+k_1 \left(\sigma^2+\zeta^2+\delta^2\right)^2\nonumber \\
		&+k_2\left(\frac{\sigma^4}{2} +\frac{\delta^4}{2}+3\sigma^2\delta^2+\zeta^4\right) +k_3\chi\left(\sigma^2-\delta^2\right)\zeta 
		-k_4\chi^4\nonumber \\
		&-\frac14\chi^4 {\rm ln}\frac{\chi^4}{\chi_0^4} +
		\frac{d}
		3\chi^4 {\rm ln}\left(\left(\frac{\left(\sigma^2-\delta^2\right)\zeta}{\sigma_0^2\zeta_0}\right)\left(\frac{\chi^3}{\chi_0^3}\right)\right),
		\label{lag_scalar}
	\end{eqnarray}	
	which includes the scalar dilaton (glueball) field $\chi$ to incorporate the broken scale invariance \cite{schechter,heide}. The second term, $\mathcal{L}_{VV}$, accounts for the self-interactions among vector mesons
	\begin{align}
		\mathcal{L}_{VV} = \frac{1}{2} \frac{\chi^2}{\chi_0^2} (m_\omega^2 \omega^2 + m_\rho^2 \rho^2 + m_\phi^2 \phi^2) + g_4 (\omega^4 + 6\omega^2\rho^2 + \rho^4 + 2\phi^4).
	\end{align}
		Lastly, the term $\mathcal{L}_{SB}$ is included to explicitly break the chiral symmetry in the model, thereby providing mass to the pseudoscalar mesons. This term is defined as 
		\begin{equation} 
			{\cal L}_{SB}=-\frac{\chi^2}{\chi_0^2}\left(h_x\sigma + 
			h_y\zeta\right),
			\label{esb_ldensity}
		\end{equation} 
		with $h_x= m^2_{\pi}f_{\pi}$ and $h_y = (\sqrt{2}m_K^2f_K-\frac{1}{\sqrt{2}}m_\pi^2f_\pi)$. An additional term, ${\cal L}_{\Delta m} = - \Delta m_s \bar \psi S \psi$, is introduced in Eq. (\ref{lag_chiral}) to fine tune the strange quark mass, with $\Delta m_s = 29$ MeV and $S \, = \, \frac{1}{3} \, \left(I - \lambda_8\sqrt{3}\right) = {\rm diag}(0,0,1)$. Within the mean-field approximation, the meson fields are replaced by their expectation values. The scalar fields: non-strange $\sigma$, strange $\zeta$, and the isovector scalar $\delta$ fields (for isospin asymmetric matter) and the vector fields: $\omega^{\mu}\rightarrow\omega\delta^{\mu 0}$, $\rho^{\mu a}\rightarrow\rho\delta^{\mu 0}\delta_{a3}$, and, $\phi^{\mu}\rightarrow\phi\delta^{\mu 0}$ have nonzero expectation values.
		The parameters of Lagrangian ($k_0,k_1,...$) and the quark-meson couplings ($g_s,g_v$), are fixed by requiring the model to reproduce key properties of QCD in the vacuum. These constraints include the masses of low-lying scalar and pseudoscalar mesons ($\pi,K, \sigma,\eta,\eta^{'}$) in vacuum. The full table of parameters of the PCQMF model is provided in Table \ref{tab:1}.
	
	\begin{table}[h]
		\scriptsize{
			\centering
			\begin{tabular}{|c|c|c|c|c|c|c|c|c|c|}
				\hline
				$k_0$           & $k_1$          & $k_2$          & $k_3$         & $k_4$         & $g_s$         & $\rm{g_v}$          & $\rm{g_4}$           & $d$          & $\rho_0$(fm$^{-3}$)                            \\ \hline
				0.2002                 & 2.3882                & -19.4995              & -4.7334              & -0.06              & 4.76               & 4               & 37.5                 & 0.002               & 0.15                                  \\ \hline
				$\sigma_0$ (MeV) & $\zeta_0$(MeV)  & $\chi_0$(MeV)   & $m_\pi$(MeV)  & $f_\pi$(MeV)  & $m_K$(MeV)    & $f_K$(MeV)     & $m_\omega$(MeV) & $m_\phi$(MeV)  & $m_\rho$( MeV)                   \\ \hline
				-93                  & -95.47              & 254.6               & 139                & 93                 & 496                & 115                 & 783                  & 1020                & 783                                   \\ \hline
				$g_{\sigma}^u$  & $g_{\sigma}^d$ & $g_{\sigma}^s$ & $g_{\zeta}^u$ & $g_{\zeta}^d$ & $g_{\zeta}^s$ & $g_{\delta}^u$ & $g_{\delta}^d$  & $g_{\delta}^s$ & $g^u_{\omega}$ \\ \hline
				$g_s/\sqrt{2}$                 & $g_s/\sqrt{2}$                & 0                   & 0                  & 0                  & $g_s$               & $g_s/\sqrt{2}$                & -$g_s/\sqrt{2}$                & 0                   &      $g_v/(2\sqrt{2})$                             \\ \hline
				$g^d_{\omega}$ & $g^s_{\omega}$ & $g^u_{\phi}$ &  $g^d_{\phi}$ & $g^s_{\phi}$  & $g^u_{\rho}$   & $g^d_{\rho}$  &   $g^s_{\rho}$  & $\Lambda_0$ (MeV)  &                             \\ \hline
				$g_v/(2\sqrt{2})$       &        0         &           0         &    0               &           $g_v/2$        &      $g_v/(2\sqrt{2})$          &    -$g_v/(2\sqrt{2})$             &          0       &               600      &                                \\ \hline
			\end{tabular}
			\caption{The list of parameters used in the present work.}
			\label{tab:1}
		}
	\end{table}

	The effective Polyakov loop potential ${\cal U}(\Phi,\bar{\Phi},T)$ in Eq. (\ref{lag_pcqmf}) takes the standard logarithmic form \cite{fuku_04}
	\begin{eqnarray}
		\frac{{\cal U}(\Phi,\bar{\Phi},T)}{T^4}&=&-\frac{a(T)}{2}\bar{\Phi}\Phi+b(T)\mathrm{ln}\big[1-6\bar{\Phi}\Phi+4(\bar{\Phi}^3+\Phi^3)-3(\bar{\Phi}\Phi)^2\big],
		\label{log}
	\end{eqnarray}
	where the temperature dependent parameters $a(T)$ and $b(T)$ are given by
	\begin{equation}\label{T}
		a(T)=a_0+a_1\bigg(\frac{T_0}{T}\bigg)+a_2\bigg(\frac{T_0}{T}\bigg)^2,\ \  b(T)=b_3\bigg(\frac{T_0}{T}\bigg)^3.
	\end{equation}
	The parameters $a_0 = 1.81$, $a_1 = -2.47$, $a_2 = 15.2$, and $b_3 = -1.75$ are determined to reproduce thermodynamic results from pure gauge lattice QCD simulations \cite{roessner}. The parameter $T_0$ represents the critical temperature and in the pure gauge sector is typically $\sim 270$ MeV \cite{fukugita}. However, when fermions are included, its value depends on the number of quark flavors $N_f$ \cite{herbst}. Additionally, the presence of these dynamical quarks influences the gluons that shape the effective Polyakov loop potential (quark backreaction). To account for dynamical quark effects on the Polyakov loop, an effective glue potential is introduced \cite{lisa}, with the mapping between glue and pure gauge temperatures defined as 
	\begin{equation}
			\frac{T_{YM} - T_{0}^{YM}}{T_{0}^{YM}} = 0.57 \frac{T_{glue} - T_{0}^{glue}}{T_{0}^{glue}},
			\qquad
			\frac{\mathcal{U}_{glue}(\Phi,\bar{\Phi},T_{glue})}{T_{glue}^4} = \frac{\mathcal{U}_{YM}(\Phi,\bar{\Phi},T_{YM})}{T_{YM}^4},
			\label{eq:glue_relations}
	\end{equation}
	where $T_{glue}$ is identified with the system temperature $T$, and $T_{0}^{YM} = T_{0}^{glue} = 200$ MeV, in this work.
	
	In the mean field approximation, the thermodynamic potential per unit volume in a grand canonical ensemble at a given temperature $T$ and chemical potential $\mu$ takes the form \cite{dj25}
	\begin{equation}
		\Omega_{\text{PCQMF}} = \Omega_{q\bar{q}} + \Omega_{\text{vac}} + {\cal U}(\Phi,\bar{\Phi},T) - \mathcal{L}_{M} - \cal V_{\text{vac}},
		\label{tpd}
	\end{equation}
	where $\Omega_{q\bar{q}}$ comprises of quark and antiquark thermal contribution and is written as
	\begin{align}
		\Omega_{q\bar{q}} = \sum_{i=u,d,s}\frac{-\gamma_i T}{(2\pi)^3}\int 
		d^3k\left\{ {\rm ln} F^{-}_i(k) +	{\rm ln} F^{+}_i(k)\right\},
	\end{align}
	with
	\begin{eqnarray}
		F^{-}_i(k)=&1+e^{-3E^-_i(k)}+3\Phi e^{-E^-_i(k)}+3\bar{\Phi}e^{-2E^-_i(k)}, \\
		F^{+}_i(k)=&1+e^{-3E^+_i(k)}+
		3\bar{\Phi} e^{-E^+_i(k)}
		+3\Phi e^{-2E^+_i(k)},
	\end{eqnarray} 
	and $\gamma_i=2$ is the spin degeneracy. The dimensionless energies are defined as $E^{\pm}_i(k) = (E_i^*(k)\pm{\mu_i}^{*})/T$ with $E_i^*(k) = \sqrt{k^2 + m_i^{*2}}$ being the effective energy of a quark of flavor $i$ and $m_i^*$ being the effective quark mass generated dynamically through its coupling to the scalar fields and is given as
	\begin{equation}
		{m_i}^{*} = -g_{\sigma}^i\sigma - g_{\zeta}^i\zeta - g_{\delta}^i\delta + \Delta m_i,
		\label{mbeff}
	\end{equation}
	where $g^i_{\sigma,\zeta,\delta}$ are the respective scalar coupling constants and $\Delta m_{u,d}=0$. Similarly, the effective chemical potential $\mu_i^*$ is shifted from its vacuum value $\mu_i$ and is defined in terms of vector fields through relation
	\begin{equation}
		{\mu_i}^{*}=\mu_i-g_\omega^i\omega-g_\phi^i\phi-g_\rho^i\rho.
		\label{mueff}
	\end{equation}
	The vector coupling constants $g^i_{\omega,\phi,\rho}$ determine the strength of the repulsive vector interactions. For asymmetric quark matter, the quark chemical potentials are related to the baryon $\mu_B$, isospin $\mu_I$, and strangeness $\mu_S$ chemical potentials through the relations $\mu_B = \frac{3}{2}(\mu_u+\mu_d),\mu_I = \frac{1}{2}(\mu_u-\mu_d)$, and $\mu_S = \frac{1}{2}(\mu_u+\mu_d-2\mu_s)$. The fermionic vacuum term, $\Omega_{\text{vac}}$, in Eq. (\ref{tpd}) is given by
	\begin{eqnarray}
		\Omega_{vac} = -2N_c \sum_{i=u,d,s}\int\frac{d^3k}{(2\pi)^3}E_i^*(k).
	\end{eqnarray}
	This integral is divergent and regularized using dimensional regularization as \cite{chatterjee2012}
	\begin{eqnarray}
		\Omega_{vac} = -\frac{N_c}{(8\pi^2)}\sum_{i=u,d,s}m_i^{*2}{\rm ln}\left(\frac{m_i^*}{\Lambda_0}\right),
	\end{eqnarray}
	where $\Lambda_0=600$ MeV is the regularization scale parameter. 
	A constant, $\cal V_{\text{vac}}$ in Eq. (\ref{tpd}) is subtracted to ensure zero vacuum energy. 
	The temperature dependence of the scalar, vector, and Polyakov fields is determined by solving the coupled equations of motion obtained by minimizing $\Omega_{\text{PCQMF}}$ with respect to each field. The scalar density $\rho_i^s$ and the vector density $\rho_i$ for each quark flavor are then calculated as 
	\begin{align}
		\rho_{i}^{s} &= \gamma_{i}N_c\int\frac{d^{3}k}{(2\pi)^{3}} 
		\frac{m_{i}^{*}}{E^{\ast}_i(k)} \Big(f_i(k)+\bar{f}_i(k)
		\Big),
		\label{rhos0}\\
		\rho_{i} &= \gamma_{i}N_c\int\frac{d^{3}k}{(2\pi)^{3}}  
		\Big(f_i(k)-\bar{f}_i(k)
		\Big),
		\label{rhov0}
	\end{align} 
	respectively. The equilibrium distribution functions for quarks $f_i(k)$ and antiquarks $\bar{f}_i(k)$, which include the effects of the Polyakov loop, are given by \cite{manisha21}
	\begin{eqnarray}
		f_{i}(k)&=&\frac{\Phi e^{-E_i^-(k)}+2\bar{\Phi} e^{-2E_i^-(k)}+e^{-3E_i^-(k)}}
		{[1+3\Phi e^{-E_i^-(k)}+3\bar{\Phi} e^{-2E_i^-(k)}+e^{-3E_i^-(k)}]} ,
		\label{distribution}
	\end{eqnarray}
	\begin{eqnarray}
		\bar{f}_{i}(k)&=&\frac{\bar{\Phi} e^{-E_i^+(k)}+2\Phi e^{-2E_i^+(k)}+e^{-3E_i^+(k)}}
		{[1+3\bar{\Phi} e_i^{-E^+(k)}+3\Phi e^{-2E_i^+(k)}+e^{-3E_i^+(k)}]}.
		\label{adistribution}
	\end{eqnarray} 
	
	\subsection{\label{anisotropy} Incorporation of Momentum-Space Anisotropy}
	To phenomenologically account for the effects of local momentum-space anisotropy, we follow the spheroidal momentum deformation proposed by Romatschke and Strickland \cite{romat}. This approach is characterized by rescaling momentum in a preferred direction and introducing a dimensionless anisotropy parameter, $\xi$, which is defined based on the components of momentum parallel ($p_{\parallel}$) and perpendicular ($\mathbf{p_{\perp}}$) to the direction of anisotropy ($\hat{\mathbf{n}}$) as
	\begin{equation}
		\xi = \frac{\langle \mathbf{p_{\perp}^2}\rangle}{2\langle p_{\parallel}^2\rangle} - 1,
	\end{equation}
	with $-1<\xi<\infty$. The parameter $\xi$ has a value of $-1$ before the collision when the whole of the system's momentum is along the longitudinal direction, implying a maximally stretched distribution along the direction of anisotropy. For values of $-1<\xi<0$, the particle distribution is stretched along the anisotropy direction, while $\xi > 0$ indicates the particle distribution is squeezed (contracted) along that direction. $\xi=0$ corresponds to an isotropic momentum distribution while $\xi=\infty$ implies that all particles are moving along the transverse direction or maximally squeezed along the anisotropy direction.
	
	For a given anisotropy direction $\hat{\mathbf{n}}$, the quark momentum, $\mathbf{k}$, is modified to $\tilde{k}$
		\begin{equation}
			\tilde{k}^2 = k^2 + \xi(\mathbf{k}\cdot\hat{\mathbf{n}})^2,
		\end{equation}
		which in turn modifies the effective single-particle energy of a quark of flavor $i$ to
		\begin{equation}
			E_{i,\text{aniso}}^*(k) = \sqrt{\tilde{k}^2 + m_i^{2}} = \sqrt{E_i^{*2}(k) + \xi k^2 \cos^2 \theta},
		\end{equation}
		where $\theta$ is the angle between $\mathbf{k}$ and the anisotropy axis. For a weakly anisotropic medium where $|\xi| < 1$, we expand this energy to linear order in $\xi$ as
		\begin{equation}
			E_{i,\text{aniso}}^*(k) \approx E_i^{*}(k) + \xi \left(\frac{ k^2 \rm{cos}^2 \theta}{2 E_i^{*}(k)}\right).
		\end{equation}
		This change to the effective energy propagates to the dimensionless energies $E^{\pm}(k) = (E_i^*(k) \pm \mu_i^*)/T$ that appear in the thermodynamic potential. The anisotropic dimensionless energy can thus be written as
		\begin{equation}
			E^{\pm}_{i,\text{aniso}}(k) = E^{\pm}_i(k) + \delta E,
		\end{equation}
		where the dimensionless correction, $\delta E$, is given by
		\begin{equation}
			\delta E = \xi\left(\frac{k^2 \rm{cos}^2 \theta}{2 E^{*}_{i}(k) T}\right).
		\end{equation}

	
	Thus, introducing momentum anisotropy into the system leads to the single particle effective energy to become direction dependent. In particular, particles moving closer to the anisotropic direction (typically the beam axis in uRHIC) acquire an additional contribution to their energy proportional to $\rm{cos}^2\theta$. This directional deformation of the effective energy induces anisotropic modification in the thermodynamic potential density of the PCQMF model. This results in anisotropic corrections to the thermodynamic and transport properties.
	The anisotropic functions are obtained by substituting $E_{i,\text{aniso}}^{\pm}(k)$ for $E_i^{\pm}(k)$ in their expressions. For any function $g(E_i^{\pm}(k))$ that depends on the dimensionless quark energy, its anisotropic counterpart $g(E_{i,\text{aniso}}^{\pm}(k))$, can be approximated to linear order in $\xi$ as
	\begin{equation}
		g(E_{i,\text{aniso}}^{\pm}(k)) \approx g(E_i^{\pm}(k)) + \frac{\partial g(E_i^{\pm}(k))}{\partial E_i^{\pm}(k)} \delta E.
		\label{aniso_function}
	\end{equation}
	We apply this linear expansion to the quark-antiquark contribution in the thermodynamic potential, $\Omega_{q\bar{q}}$, gets modified as
	\begin{equation}
		\Omega_{q\bar{q}, \text{aniso}} = \sum_{i=u,d,s}\frac{-\gamma_i T}{(2\pi)^3}\int d^3k\left\{ {\rm ln} F^{-}_{i,\text{aniso}}(k) +	{\rm ln} F^{+}_{i,\text{aniso}}(k) \right\}.
		\label{omega_aniso}
	\end{equation}
	Using Eq. (\ref{aniso_function}) for ${\rm ln} F^{\pm}_{i,\text{aniso}}(k)$, the Eq. (\ref{omega_aniso}) can be written as
	\begin{equation}
		\Omega_{q\bar{q}, \text{aniso}} = \Omega_{q\bar{q}} + \xi \left( \sum_{i=u,d,s}\frac{\gamma_iN_c}{12\pi^2E_i^*(k)} \int k^4 dk [f_i(k) + \bar{f}_i(k)] \right).
	\end{equation}
	Similarly, the anisotropic scalar and vector densities can be identified by differentiating the above equation with respect to $\sigma$ and $\omega$, respectively, as
	\begin{equation}
		\rho_{i,\text{aniso}}^s = \rho_i^s - \xi\left(  \frac{\gamma_i N_c}{12\pi^2 T}\int k^4 dk \frac{m_i^*}{E_i^{*2}(k)}\left[ A_i(k) + B_i(k) \right]\right),
	\end{equation}
	\begin{equation}
		\rho_{i,\text{aniso}} = \rho_i + \xi\left(  \frac{\gamma_i N_c}{12\pi^2 T}\int \frac{k^4 dk}{E_i^*(k)} \left[ A_i(k) - B_i(k) \right]\right),
	\end{equation}
	where the terms
	\begin{equation}
		A_i(k) = \frac{\left(\Phi e^{-E_i^-(k)} + 4\bar{\Phi}e^{-2E_i^-(k)} + 3e^{-3E_i^-(k)}\right) - 3f_i(k)\left(\Phi e^{-E_i^-(k)} + 2\bar{\Phi}e^{-2E_i^-(k)} + e^{-3E_i^-(k)}\right)}{F_i^-(k)}
	\end{equation}
	and
	\begin{equation}
		B_i(k) = \frac{\left(\bar{\Phi} e^{-E_i^+(k)} + 4{\Phi}e^{-2E_i^+(k)} + 3e^{-3E_i^+(k)}\right) - 3\bar{f}_i(k)\left(\bar{\Phi} e^{-E_i^+(k)} + 2{\Phi}e^{-2E_i^+(k)} + e^{-3E_i^+(k)}\right)}{F_i^+(k)}.
	\end{equation}	
	The anisotropic distribution functions, $f_{i,\text{aniso}}(k)$ and $\bar{f}_{i,\text{aniso}}(k)$ are then approximated as 
	\begin{equation}
		f_{i,\text{aniso}}(k) \approx f_i(k) + \left(\frac{\partial f_i(k)}{\partial E_i^{-}(k)}\right) \delta E \equiv f_i(k) + \delta f_{i}(k),
	\end{equation}
	\begin{equation}
		\bar{f}_{i,\text{aniso}}(k) \approx \bar{f}_i(k) + \left(\frac{\partial \bar{f}_i(k)}{\partial E_i^{+}(k)}\right) \delta E \equiv \bar{f}_i(k) + \delta \bar{f}_{i}(k),
	\end{equation}
	which can be split into isotropic and anisotropic parts as
		\begin{equation}
			f_{i,\text{aniso}}(k) = f_i(k) - \xi \left( \frac{k^2 cos^2 \theta}{2 E_i^*(k) T} \left[ f_i(k)(1-3f_i(k)) + X_i(k) \right] \right),
			\label{faniso}
		\end{equation}
		\begin{equation}
			\bar{f}_{i,\text{aniso}}(k) = \bar{f}_i(k) - \xi \left( \frac{k^2 cos^2 \theta}{2 E_i^*(k) T} \left[ \bar{f}_i(k)(1-3\bar{f}_i(k)) + Y_i(k) \right] \right),
			\label{fbaraniso}
		\end{equation}
		with
		\begin{equation}
			X_i(k) = \frac{2\left(\bar{\Phi}e^{-2E_i^-(k)} + e^{-3 E_i^-(k)}\right)}{F_i^-(k)},
		\qquad
			Y_i(k) = \frac{2\left({\Phi}e^{-2E_i^+(k)} + e^{-3 E_i^+(k)}\right)}{F_i^+(k)}.
		\end{equation}	
	
	\subsection{\label{transport} Transport coefficients in anisotropic PCQMF model}
	
	The transport coefficients can be derived from the RBE, which describes the evolution of the single particle distribution function in phase space. For systems close to the equilibrium in presence of external field, one can use the RTA to simplify the complex collision term, and the RBE is then written as \cite{groot}
	\begin{equation} 
			k^{\mu}\partial_{\mu}f(x, k) + qF^{\alpha \beta}k_{\beta}\frac{\partial}{\partial k^{\alpha}}f(x,k)= C[f(x, k)] = - \frac{k^{\mu}u_{\mu}}{\tau} (f(x, k) - f_0(x, k)) = - \frac{k^{\mu}u_{\mu}}{\tau} \delta f,
	\end{equation}
	where $q$ is the charge of a quark and $F^{\alpha \beta}$ represent the electromagnetic field strength tensor. The term $u_{\mu}$ is the fluid four-velocity, $\tau$ is the relaxation time for the particle species, and $\delta f = f(x,k) - f_0(x,k)$ is the deviation from the local equilibrium distribution $f_0(x,k)$ given in Eq. (\ref{distribution}). The relaxation time $\tau$ is given by \cite{hosoya}
	\begin{equation}
		\tau = \frac{1}{5.1T\alpha_S^2{\rm ln}(1/\alpha_S)(1+0.12(2N_f+1))},
		\label{tau}
	\end{equation}
	where $\alpha_s(T,\mu)$ is the standard two-loop running coupling written as \cite{bannur,zhu}
	\begin{equation}
		\alpha_S(T,\mu)=\frac{6\pi}{(33-2N_f){\rm ln}\left(\frac{T}{\Lambda_T}\sqrt{1+(\frac{\mu}{\pi T})^2}\right)}\left(1-\frac{3(153-19N_f)}{(33-2N_f)^2}\frac{{\rm ln}\left(2{\rm ln}\frac{T}{\Lambda_T}\sqrt{1+(\frac{\mu}{\pi T})^2}\right)}{{\rm ln}\left(\frac{T}{\Lambda_T}\sqrt{1+(\frac{\mu}{\pi T})^2}\right)}\right),
		\label{alpha}
	\end{equation}
	with $\Lambda_T$ being the QCD scale-fixing parameter originating from the lowest nonzero Matsubara modes \cite{vuorinen} and is fixed at $\Lambda_T=70$ MeV \cite{zhu}. We assume $\tau$ to remain isotropic even in the presence of momentum anisotropy in the distribution functions. This is a simplifying approximation that allows us to isolate the direct effect of the anisotropic particle distribution on the transport coefficients. While a fully self-consistent treatment would involve an anisotropic relaxation time, this approach is standard in many initial phenomenological studies \cite{zhang21,lata17,srivastava}.
	
	The transport coefficients quantify the linear response of a system to small gradients or external fields. They are extracted from the dissipative parts of the energy-momentum tensor $T^{\mu\nu}$ and conserved current densities such as electric current $J^{\mu}$. The general procedure involves measuring small perturbations in the local equilibrium distribution function $f_0$ as $\delta f = f -f_0$ using the RBE within RTA, in the presence of appropriate gradients, e.g., velocity gradient for shear and bulk viscosities, or electric field for conductivity. $\delta f$ is then substituted into the expressions of $T^{\mu\nu}$ or $J^{\mu}$, and the transport coefficients quantify the linear response of a system by relating dissipative fluxes to their associated thermodynamic gradients. The detailed derivations can be found in Refs. \cite{hosoya,plumari,chakraborty}. The general expressions in an isotropic medium are given by \cite{saha2018,islam}
	\begin{eqnarray}
		\eta&=&\frac{2N_{c}}{15T}\sum_{i=u,d,s}\int\frac{{\rm d}^3k}{(2\pi)^3}\tau\left(\frac{k^{2}}{E_i^{*}}\right)^{2}[f_{i}(1-f_{i})+\bar{f}_{i}(1-\bar{f}_{i})],
		\label{eta}
	\end{eqnarray}
	\begin{eqnarray}
		\zeta_b&=&\frac{2N_{c}}{T}\sum_{i=u,d,s}\int\frac{{\rm d}^3k}{(2\pi)^3}\tau\frac{1}{E_i^{*2}}\left[\left(\frac{1}{3}-c_{s}^{2}\right)k^{2}-c_{s}^{2}m_i^{*2}+c_{s}^{2}m_i^{*}T\frac{dm_i^{*}}{dT}\right]^{2} \nonumber \\ 
		&&\left[f_{i}(1-f_{i})+\bar{f}_{i}(1-\bar{f}_{i})\right], 
		\label{zeta}
	\end{eqnarray}
	\begin{eqnarray}
		\sigma_{el}&=&\frac{2N_{c}}{3T}\sum_{i=u,d,s}e_i^{2}\int\frac{{\rm d}^3k}{(2\pi)^3}\tau\left(\frac{k}{E_i^{*}}\right)^{2}[f_{i}(1-f_{i})+\bar{f}_{i}(1-\bar{f}_{i})],
		\label{sigma}
	\end{eqnarray}
	The term $c_s^2$ is the squared speed of sound and characterizes the medium's response to pressure disturbances. At constant entropy, it is defined as
	\begin{equation}
		c_s^2 = \left(\frac{\partial p}{\partial \epsilon}\right)_{s} = \frac{s}{c_{v}},
		\label{sound_eq}
	\end{equation}
	where the pressure $p$, energy density $\epsilon$, entropy density $s$, and specific heat at constant volume $c_v$ in an anisotropic medium are expressed as
	\begin{equation}
		p = -\Omega_{\text{aniso}},
		\label{pressure}
	\end{equation}
	\begin{equation}
		\epsilon = \Omega_{\text{aniso}}+\sum_{i} {\mu_i}^{*} \rho_{i,\text{aniso}}+Ts,
		\label{energy}
	\end{equation}
	\begin{equation}
		s= -\frac{\partial\Omega_{\text{aniso}}}{\partial T},
		\label{entropy}
	\end{equation}
	\begin{equation}
		c_{v} = \left(\frac{\partial \epsilon}{\partial T}\right)_V,
		\label{heat}
	\end{equation}
	where 
		\begin{equation}
			\Omega_{\text{aniso}} = \Omega_{q\bar{q}, \text{aniso}} + \Omega_{vac} + {\cal U}(\Phi,\bar{\Phi},T) - \mathcal{L}_{M} - \cal V_{\text{vac}}.
	\end{equation}
	To study the effect of momentum anisotropy in the transport coefficients, the equilibrium distribution function $f_i$ (Eq. (\ref{distribution})) is replaced with $f_{i,\text{aniso}}$ (Eq. (\ref{faniso})) in the above expressions. The resulting (to linear order in $\xi$) anisotropic transport coefficients are obtained as
		\begin{eqnarray}
			\eta_{\text{aniso}} &=& \eta - \xi \frac{N_c \tau}{90 T^2 \pi^2} \sum_{i=u,d,s}\int dk \frac{k^8}{E_i^{*3}(k)}\bigg( f_i(1-f_i) + \bar{f}_i(1-\bar{f}_i) -4(f_i^2 + \bar{f}_i^2) \nonumber \\  &+& X_i(k)(1-2f_i) + Y_i(k)(1-2\bar{f}_i)\bigg),
			\label{eta_a}
		\end{eqnarray}
		\begin{eqnarray}
			\zeta_{b,\text{aniso}} &=& \zeta_b - \xi \frac{N_c \tau}{6 T^2 \pi^2} \sum_{i=u,d,s}\int dk \frac{k^4}{E_i^{*3}(k)}\left[\left(\frac{1}{3}-c_{s}^{2}\right)k^{2}-c_{s}^{2}m_i^{*2}+c_{s}^{2}m_i^{*}T\frac{dm_i^{*}}{dT}\right]^{2} \nonumber \\  
			&&\bigg( f_i(1-f_i) + \bar{f}_i(1-\bar{f}_i) -4(f_i^2 + \bar{f}_i^2) + X_i(k)(1-2f_i) + Y_i(k)(1-2\bar{f}_i)\bigg),
			\label{zeta_a}
		\end{eqnarray}
		\begin{eqnarray}
			\sigma_{el,\text{aniso}} &=& \sigma_{el} - \xi \frac{N_c \tau}{18 T^2 \pi^2} \sum_{i=u,d,s} e_i^2 \int dk \frac{k^6}{E_i^{*3}(k)}\bigg( f_i(1-f_i) + \bar{f}_i(1-\bar{f}_i) -4(f_i^2 + \bar{f}_i^2) \nonumber \\ &+& X_i(k)(1-2f_i) + Y_i(k)(1-2\bar{f}_i)\bigg).
			\label{sigma_a}
	\end{eqnarray}
	In the isotropic limit ($\xi\rightarrow0$), all anisotropic corrections in the above equations vanish and one recovers their standard expressions in Eqs. (\ref{eta}) - (\ref{sigma}). The anisotropic terms are proportional to $\xi$ and introduce angular integrals (involving $\rm{cos}^2\theta$) which result in enhancing or suppressing contributions from longitudinal modes, depending on the sign of $\xi$. Although the overall form of the integrals remains similar to the isotropic case, the anisotropic components carry higher powers of momentum, encoding their sensitivity to directional flow and gradients in the medium. This decomposition of transport coefficients into isotropic and anisotropic parts enables their systematic numerical analysis.

	\section{\label{results} Results}
	
	In this section, we present and discuss the impact of momentum-space anisotropy, characterized by the parameter $\xi$, on the chiral and deconfinement order parameters, various thermodynamic observables, and key transport coefficients of strongly interacting QCD matter within the PCQMF model. The analysis is performed for two distinct scenarios: one at vanishing baryon chemical potential ($\mu_B = 0$ MeV) and another at a finite baryon chemical potential ($\mu_B = 400$ MeV). The results are presented for different values of the anisotropy parameter: the isotropic case $\xi=0.0$, oblate (squeezed) distributions with $\xi=0.2$ and $\xi=0.4$, and a prolate (stretched) case with $\xi=-0.2$. 
	
	\begin{figure}
		\centering
		\begin{minipage}[c]{0.98\textwidth}
			(a)\includegraphics[width=7.4cm]{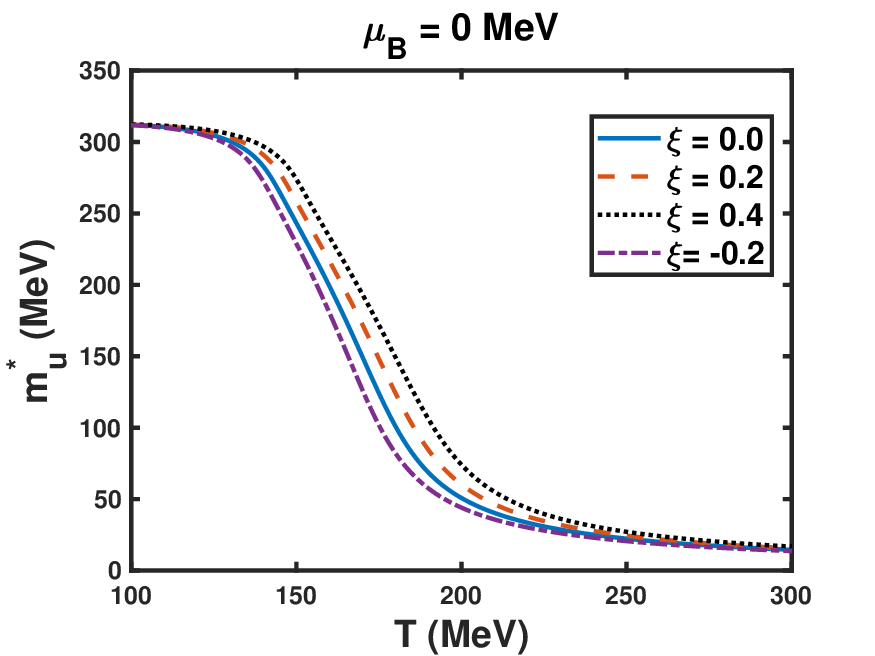}
			\hspace{0.03cm}
			(b)\includegraphics[width=7.4cm]{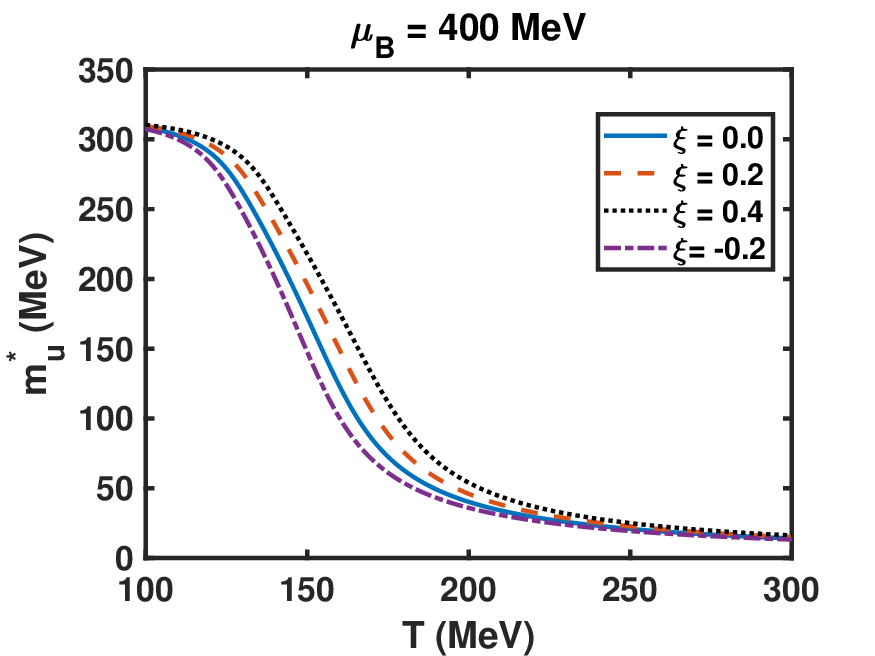}
			\hspace{0.03cm}	
			(c)\includegraphics[width=7.4cm]{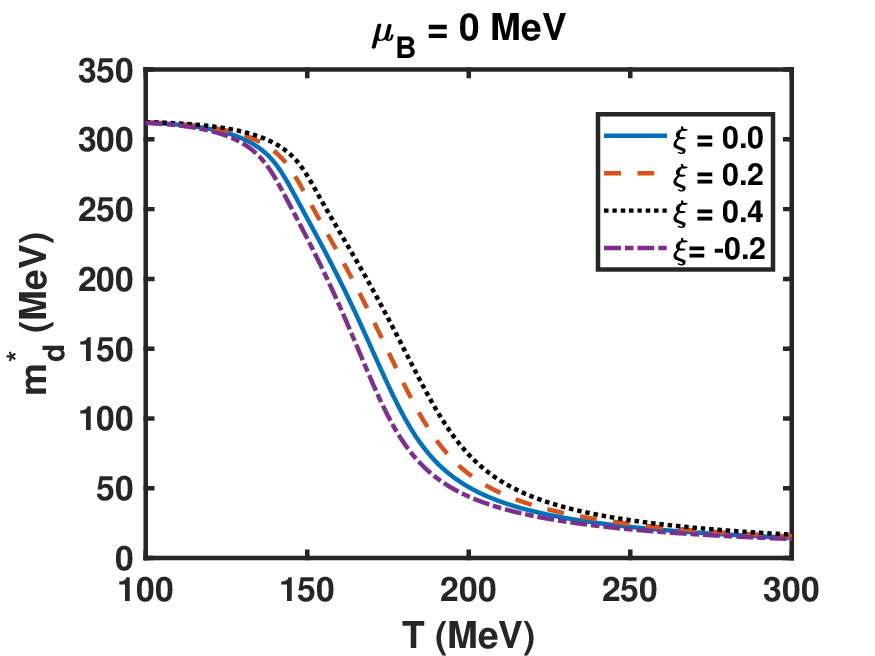}
			\hspace{0.03cm}
			(d)\includegraphics[width=7.4cm]{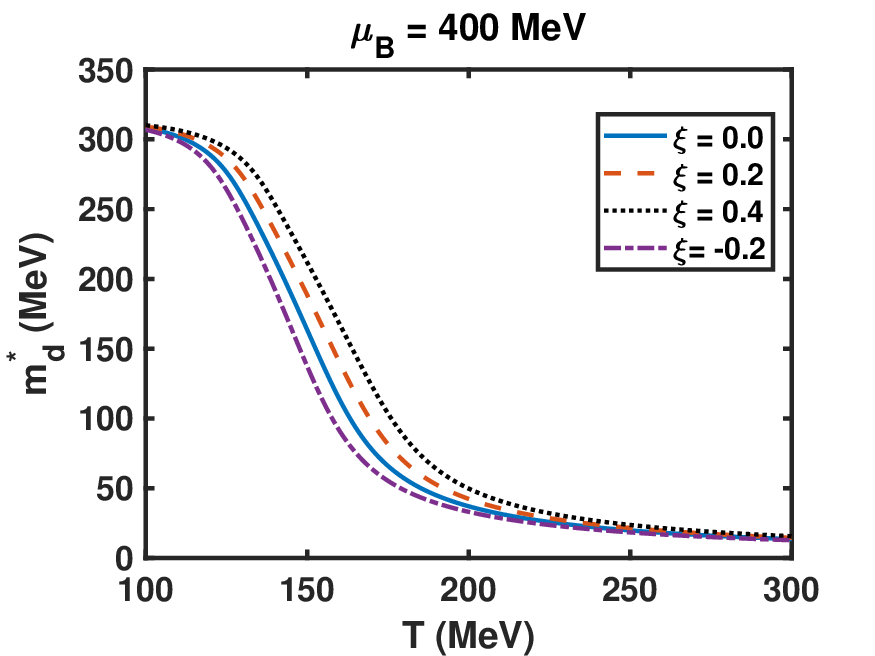}
			\hspace{0.03cm}	
			(e)\includegraphics[width=7.4cm]{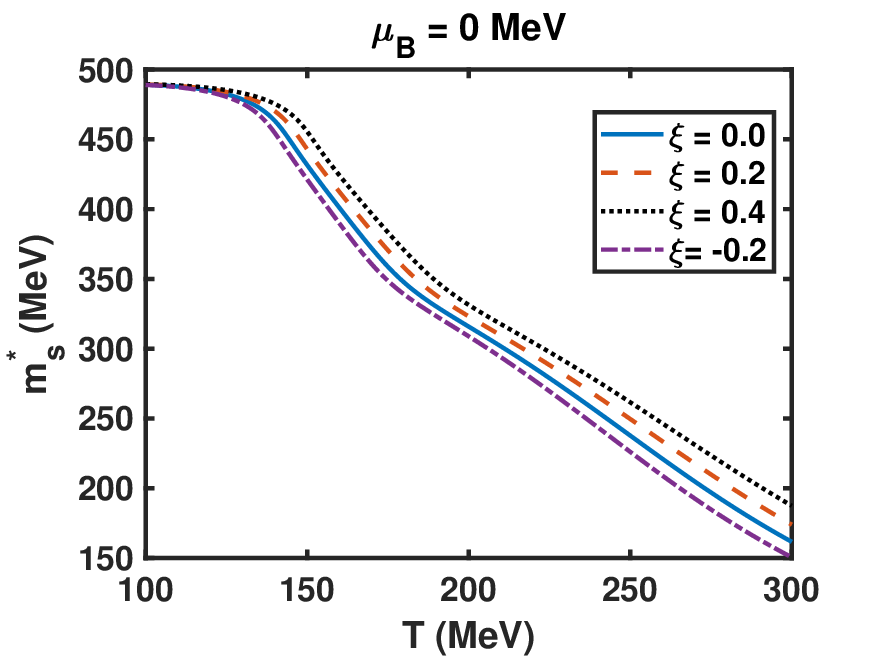}
			\hspace{0.03cm}
			(f)\includegraphics[width=7.4cm]{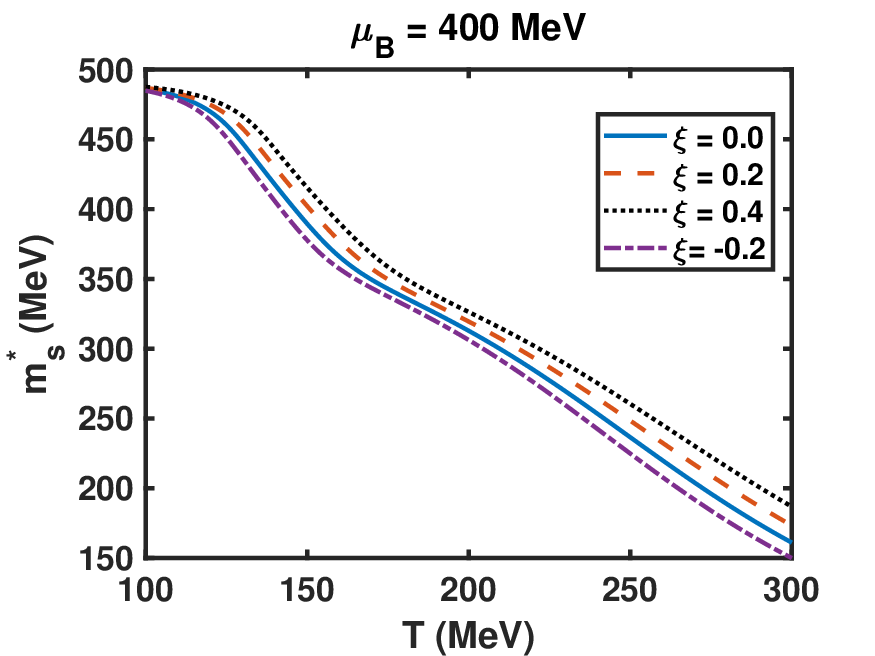}
			\hspace{0.03cm}
		\end{minipage}
		\caption{\label{fig_mass} The effective quark masses $m_u^*,m_d^*,$ and $m_s^*$ plotted as a function of temperature $T$ for anisotropy parameter $\xi=0$, 0.2, 0.4, and $-0.2$, at baryon chemical potential $\mu_B = 0$ MeV [in subplots (a), (c), and (e)], and baryon chemical potential $\mu_B = 400$ MeV, isospin chemical potential $\mu_I = -30$ MeV, and strangeness chemical potential $\mu_S = 125$ MeV [in subplots (b),(d), and (f)].}
	\end{figure}	
	
	We begin by examining the effect of momentum anisotropy on the effective quark masses, which are directly linked to the chiral dynamics of the system. Fig. \ref{fig_mass} shows the effective quark masses $m_u^*$, $m_d^*$, and $m_s^*$ as a function of temperature $T$. The left panels [(a), (c), (e)] correspond to $\mu_B = 0$ MeV, while the right panels [(b), (d), (f)] are for a system at $\mu_B = 400$ MeV, $\mu_I = -30$ MeV, and $\mu_S = 125$ MeV. For the isotropic case ($\xi = 0$), all quark masses exhibit a significant reduction with increasing temperature, reflecting the gradual restoration of chiral symmetry. When anisotropy is introduced, this behavior is modified. For positive values of $\xi$ (oblate), the quark masses decrease slowly, suggesting that a squeezed momentum distribution delays chiral symmetry restoration. Conversely, a negative value of $\xi$ (prolate) results in a faster decrease of the quark masses, indicating that the chiral transition is triggered earlier. This behavior can be understood from the modification of single-particle energy due to anisotropy. An oblate distribution enhances the effective thermal energy of the quarks, thereby strengthening the transition. Similar observations were reported in the two-flavor NJL model \cite{zhang22}. This trend persists at a finite baryon chemical potential of $\mu_B = 	400$ MeV, although the overall transition occurs at a lower temperature.
	
	\begin{figure}
		\centering
		\begin{minipage}[c]{0.98\textwidth}
			(a)\includegraphics[width=7.4cm]{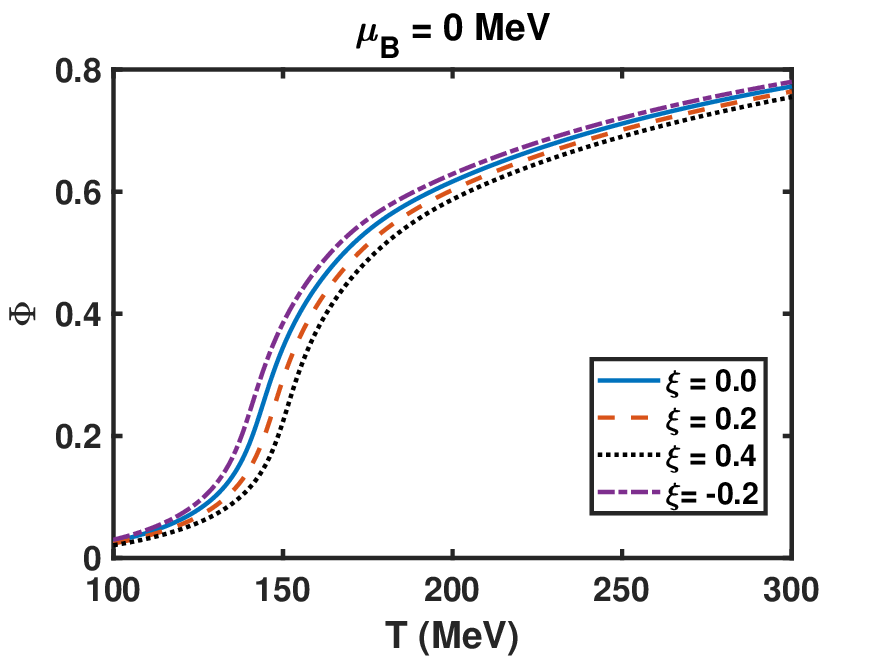}
			\hspace{0.03cm}
			(b)\includegraphics[width=7.4cm]{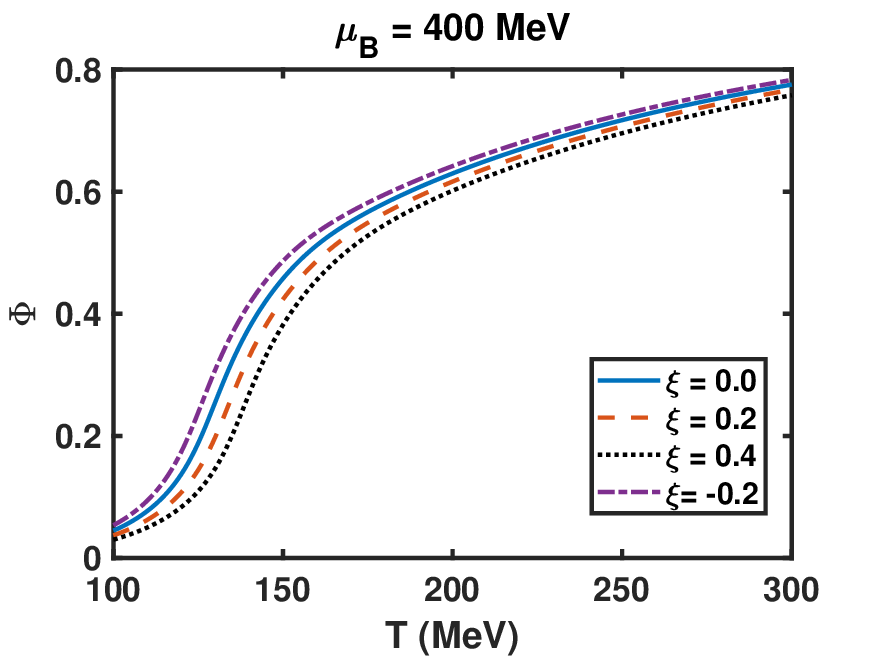}
			\hspace{0.03cm}	
			(c)\includegraphics[width=7.4cm]{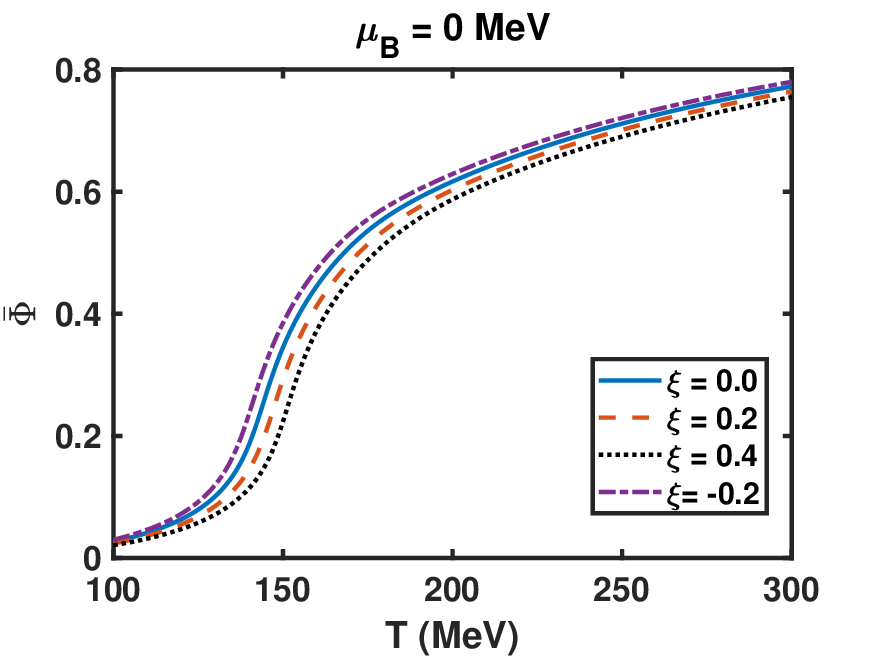}
			\hspace{0.03cm}
			(d)\includegraphics[width=7.4cm]{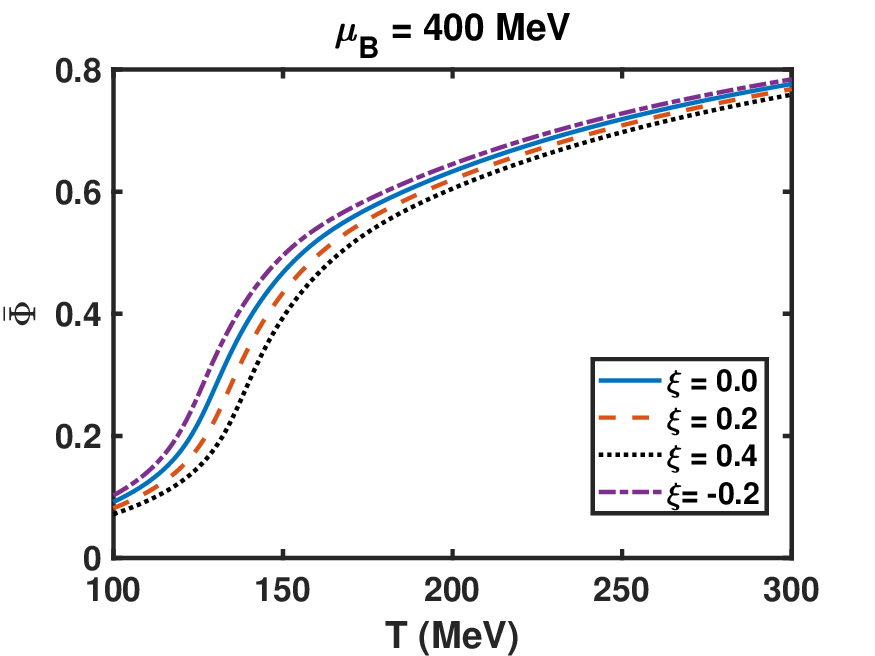}
			\hspace{0.03cm}	
		\end{minipage}
		\caption{\label{fig_poly} The Polyakov fields $\Phi$ and $\bar{\Phi}$ are plotted as a function of temperature $T$ for anisotropy parameter $\xi=0$, 0.2, 0.4, and $-0.2$, at baryon chemical potential $\mu_B = 0$ MeV [in subplots (a) and (c)], and baryon chemical potential $\mu_B = 400$ MeV, isospin chemical potential $\mu_I = -30$ MeV, and strangeness chemical potential $\mu_S = 125$ MeV [in subplots (b) and (d)].}
	\end{figure}
	
	Next, we investigate the temperature dependence of Polyakov loop fields, $\Phi$ and $\bar{\Phi}$ for various values of $\xi$ at $\mu_B=0$ MeV [(a), (c)] and $\mu_B=400$ MeV [(b), (d)] in Fig. \ref{fig_poly}. In the isotropic case, both $\Phi$ and $\bar{\Phi}$ transition from values close to zero in the low-temperature (confined) phase to values approaching unity in the high-temperature (deconfined) phase.
	A positive $\xi$ leads to a slight suppression of $\Phi$ and $\bar{\Phi}$ at intermediate temperatures, while a negative $\xi$ causes a minor enhancement. This may indicate that the deconfinement of quarks gets delayed for a system with a squeezed momentum distribution compared to the isotropic case. Increasing the baryon chemical potential to $\mu_B=400$ MeV leads to unequal enhancement in $\Phi$ and $\bar{\Phi}$ at lower temperatures. The effect of $\xi$ is similar to the case of zero chemical potential.
	
\begin{figure}
		\centering
		\begin{minipage}[c]{0.98\textwidth}
			(a)\includegraphics[width=7.4cm]{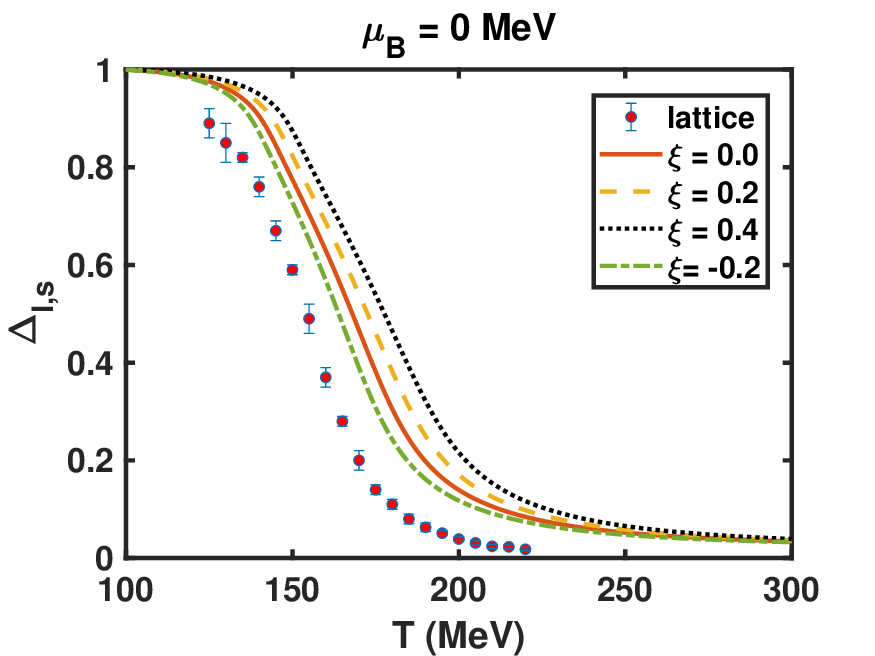}
			\hspace{0.03cm}
			(b)\includegraphics[width=7.4cm]{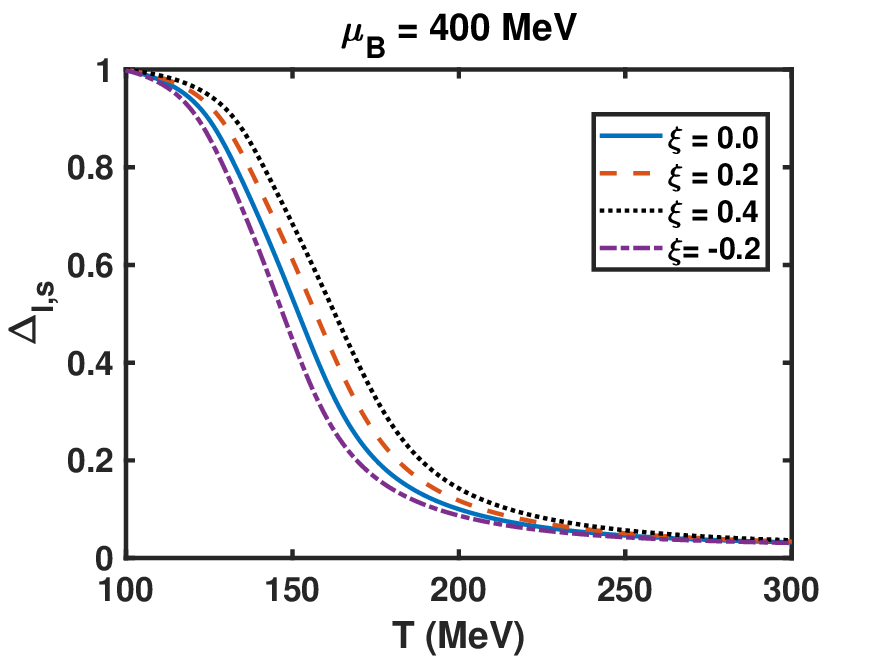}
			\hspace{0.03cm}	
		\end{minipage}
		\caption{\label{condensate} The subtracted chiral condensate $\Delta_{l,s}$ is plotted as a function of temperature $T$ with $\mu_B = 0$ MeV (a) and $\mu_B = 400$ MeV (b) at anisotropy parameter $\xi=0$, 0.2, 0.4, and $-0.2$.}
	\end{figure}

	To more precisely quantify the chiral phase transition, we compute the subtracted chiral condensate, $\Delta_{l,s}$, defined as \cite{baza2009}
	\begin{equation}
		\Delta_{l,s}(T) = \frac{\langle \bar{\psi}\psi\rangle_{l,T}-\frac{\hat{m}_l}{\hat{m}_s}\langle\bar{\psi}\psi\rangle_{s,T}}{\langle \bar{\psi}\psi\rangle_{l,0}-\frac{\hat{m}_l}{\hat{m}_s}\langle\bar{\psi}\psi\rangle_{s,0}},
	\end{equation}
	where $\hat{m}_l$ and $\hat{m}s$ denote the bare masses of light and strange quarks, respectively, and are directly related to the symmetry breaking parameters $h_x$ and $h_y$ appearing in Eq. (\ref{esb_ldensity}) \cite{schaefer}. Additionally, in the context of the PCQMF model, the light quark condensate $\langle \bar{\psi}\psi\rangle_{l,T}$ and strange quark condensate $\langle \bar{\psi}\psi\rangle_{s,T}$ can be replaced with the non-strange $\sigma$ and strange $\zeta$ field, respectively. As a result, we can write the subtracted chiral condensate as
	\begin{equation}
		\Delta_{l,s}(T) = \frac{\sigma-\frac{h_x}{h_y}\zeta}{\sigma_0-\frac{h_x}{h_y}\zeta_0},
	\end{equation}
	where $\sigma_0$ and $\zeta_0$ denote the vacuum values of the scalar fields. This observable provides a clear signal for the chiral crossover. In Fig. \ref{condensate}, we plot $\Delta_{l,s}$ as a function of temperature at various values of $\xi$. The left panel shows the results at $\mu_B=0$ MeV, which are in good qualitative agreement with lattice QCD data from Ref. \cite{borsanyi2010}.
	Positive values of $\xi$ shift the transition curve to higher temperatures, while a negative $\xi$ shifts it to lower temperatures. This confirms that a prolate momentum distribution promotes an earlier chiral transition, whereas an oblate distribution delays it. At $\mu_B=400$ MeV (right panel), the transition temperature is lowered, as expected, and the separation between the curves for different $\xi$ values remains significant, highlighting the persistent role of anisotropy at finite density.
	
\begin{figure}
		\centering
		\begin{minipage}[c]{0.98\textwidth}
			(a)\includegraphics[width=7.4cm]{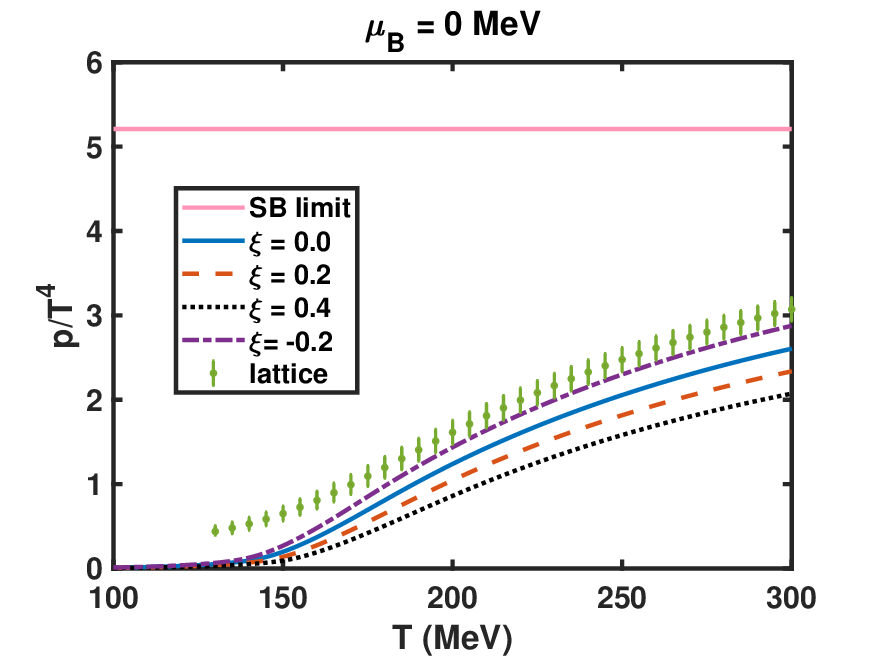}
			\hspace{0.03cm}
			(b)\includegraphics[width=7.4cm]{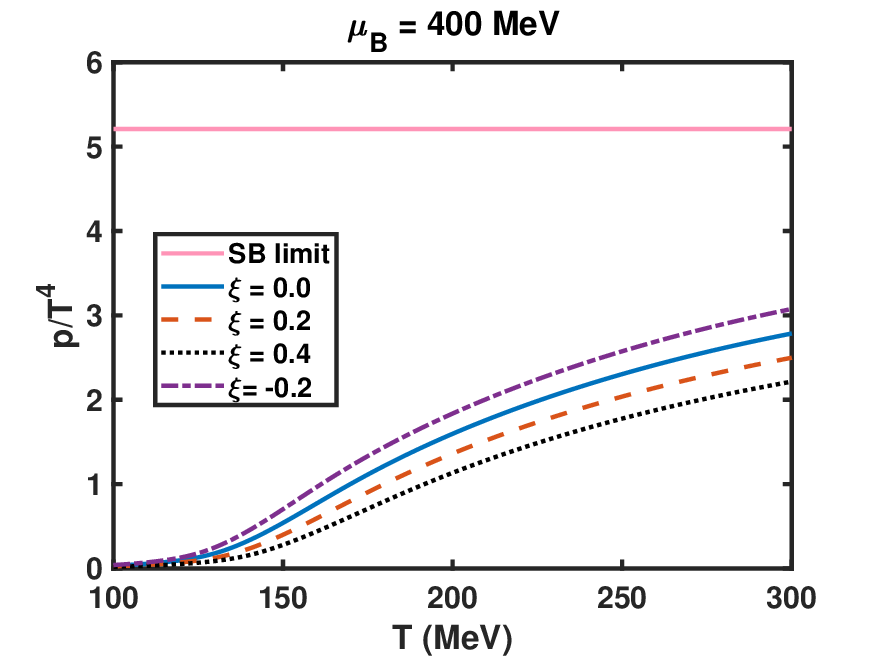}
			\hspace{0.03cm}	
			(c)\includegraphics[width=7.4cm]{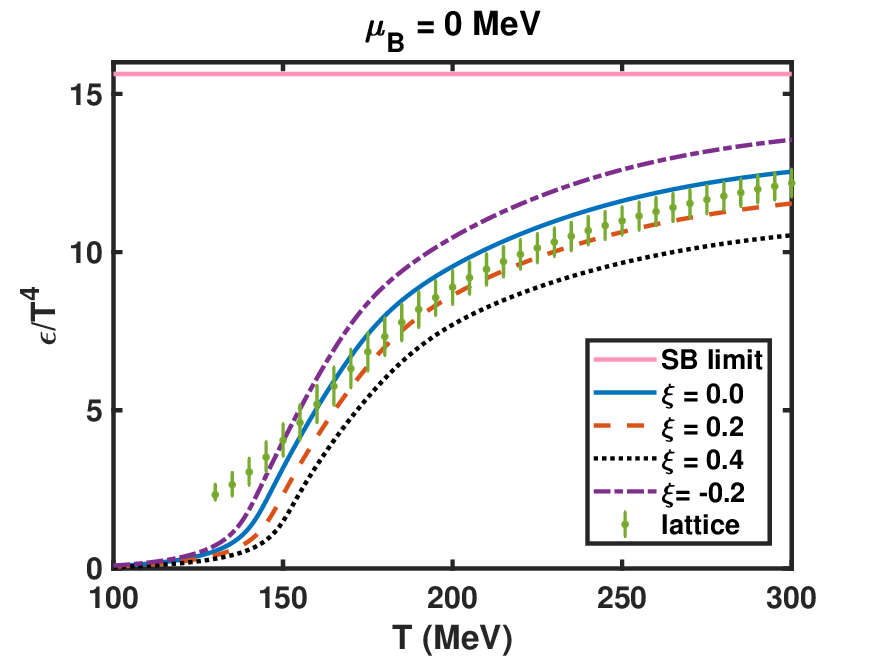}
			\hspace{0.03cm}
			(d)\includegraphics[width=7.4cm]{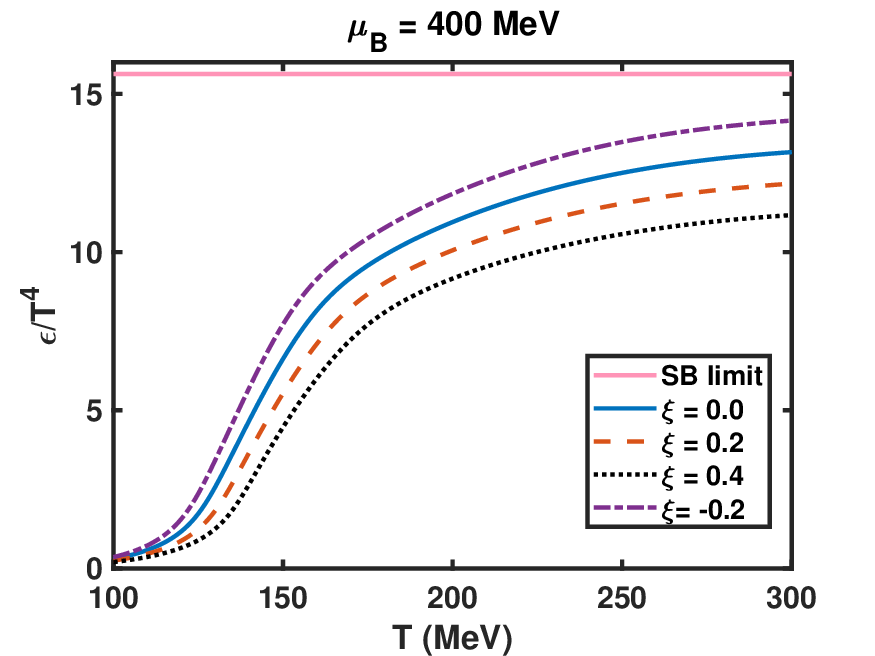}
			\hspace{0.03cm}	
			(e)\includegraphics[width=7.4cm]{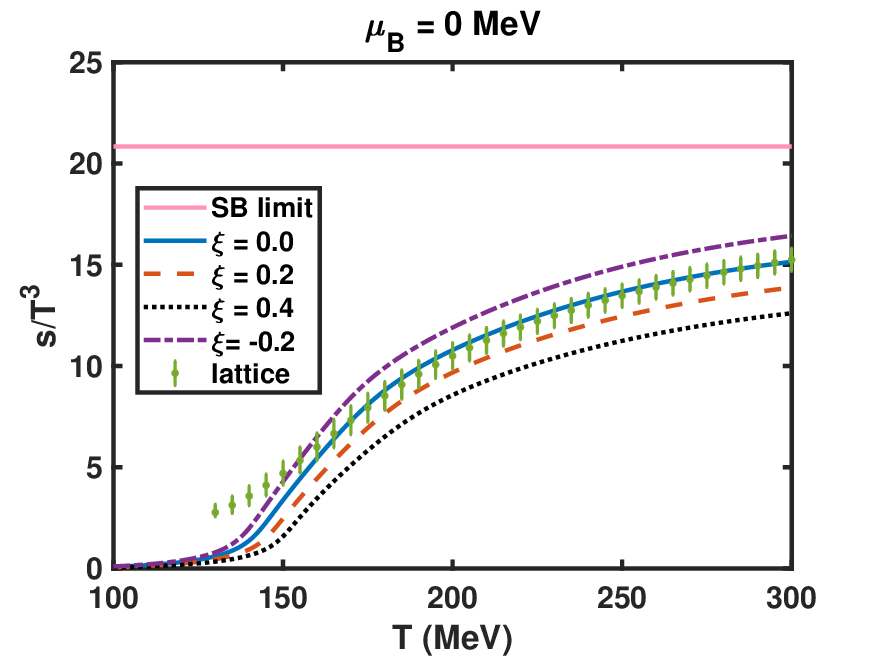}
			\hspace{0.03cm}
			(f)\includegraphics[width=7.4cm]{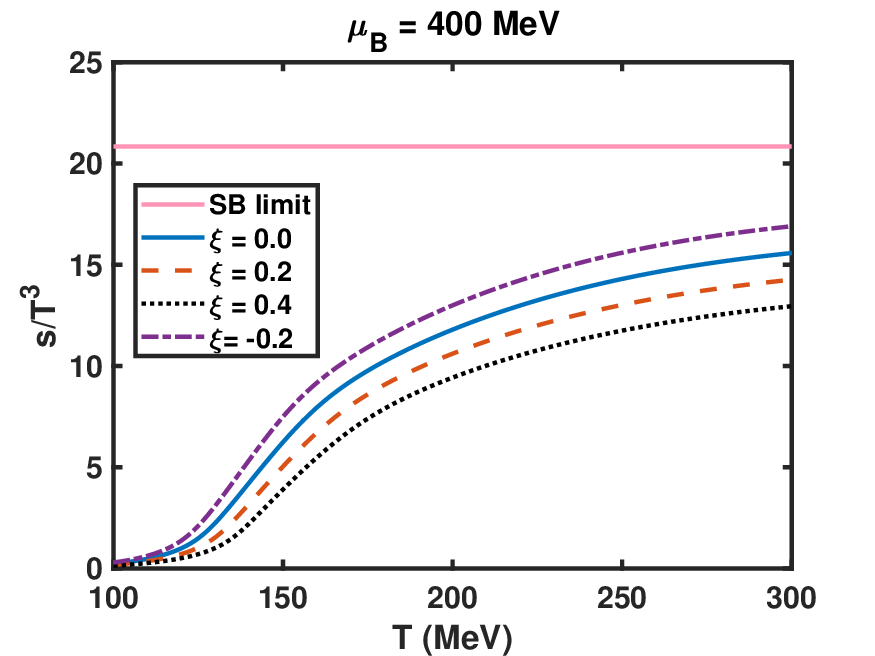}
			\hspace{0.03cm}
		\end{minipage}
		\caption{\label{thermo} The scaled pressure $p/T^4$, scaled energy density $\epsilon/T^4$, and scaled entropy density $s/T^3$ as a function of temperature $T$ for various values of anisotropy parameter $\xi$. Subplots (a), (c), and (e) correspond to a system at $\mu_B = 0$ MeV, where the results are compared with lattice QCD data from the HotQCD \cite{hot2} and Wuppertal-Budapest (WB) \cite{borsanyi} collaborations. Subplots (b), (d), and (f) are for a system with $\mu_B=400$ MeV, $\mu_I = -30$ MeV, $\mu_S=125$ MeV.}
	\end{figure}

	The impact of momentum anisotropy on the bulk thermodynamic properties of the quark matter is presented in Fig. \ref{thermo}. Here, we plot the scaled pressure $p/T^4$, scaled energy density $\epsilon/T^4$, and scaled entropy density $s/T^3$ as functions of temperature. The left panels [(a), (c), (e)] at $\mu_B=0$ are compared with lattice data. For all cases, these quantities remain small in the confined phase and rise with temperature, finally approaching the Stefan-Boltzmann (SB) limit at very high $T$. Our results are in good qualitative agreement with the lattice data. A key finding is that the equation of state is directly modified by momentum anisotropy. Positive $\xi$ values lead to a suppression in pressure, energy density, and entropy density at a given temperature, while negative $\xi$ values lead to their enhancement. This implies that a system with a stretched momentum distribution is thermodynamically more active than an isotropic one. At finite baryon chemical potential $\mu_B = 400$ MeV, these thermodynamic quantities are larger, particularly at lower temperatures, due to the finite quark density, but the qualitative impact of $\xi$ remains the same.
	
	\begin{figure}
		\centering
		\begin{minipage}[c]{0.98\textwidth}
			(a)\includegraphics[width=7.4cm]{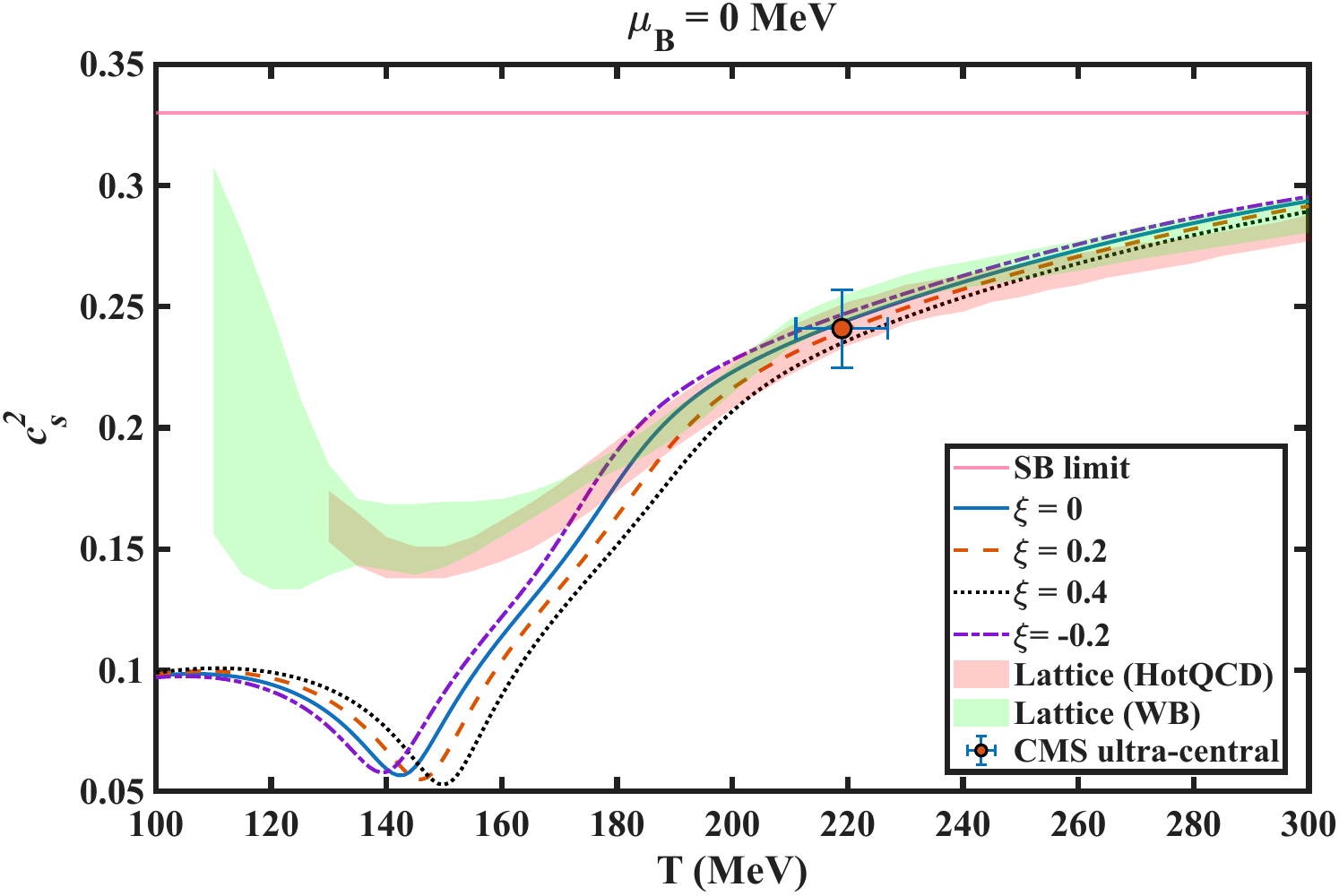}
			\hspace{0.03cm}
			(b)\includegraphics[width=7.4cm]{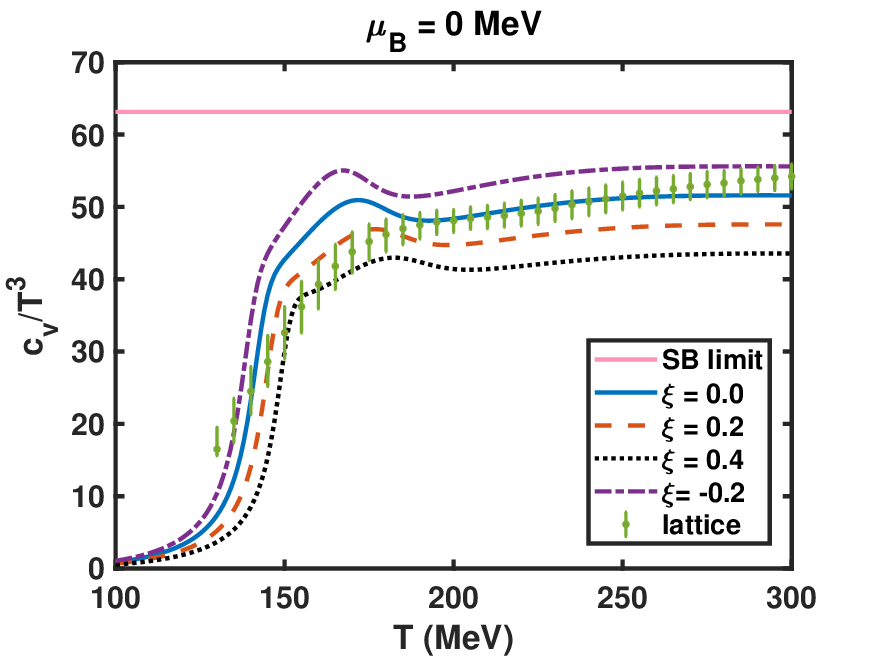}
			\hspace{0.03cm}	
			(c)\includegraphics[width=7.4cm]{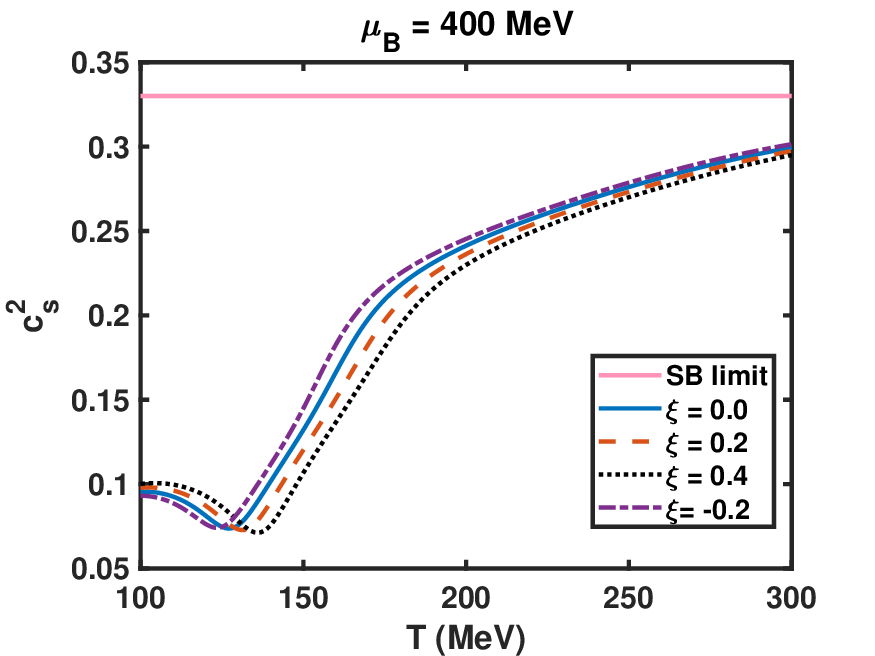}
			\hspace{0.03cm}
			(d)\includegraphics[width=7.4cm]{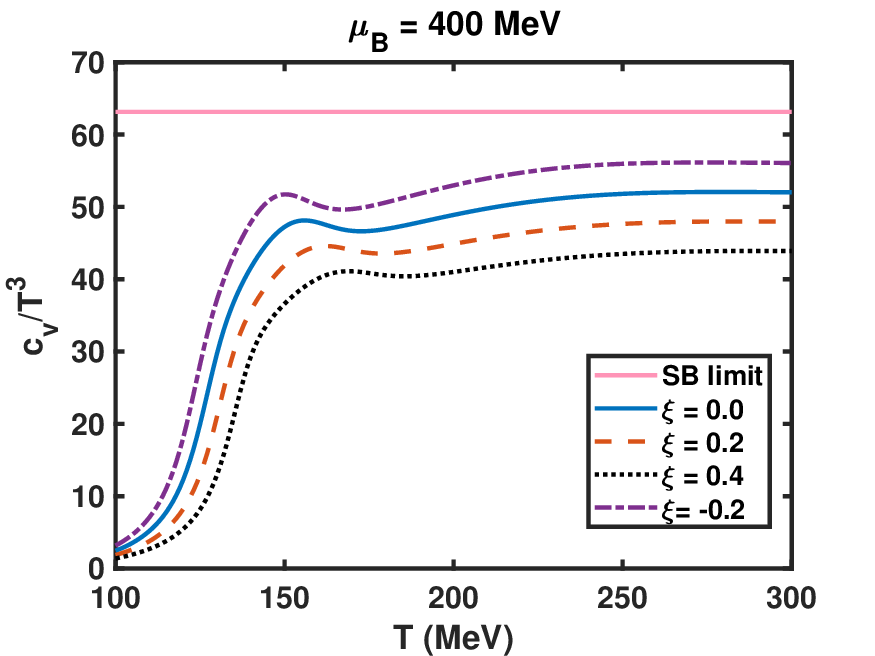}
			\hspace{0.03cm}	
		\end{minipage}
		\caption{\label{sound} The temperature dependence of speed of sound squared $c_s^2$ and scaled specific heat $c_v/T^3$ for various values of anisotropy parameter $\xi=0$, 0.2, 0.4, and $-0.2$. The top row [subplots (a) and (b)] corresponds to a system at $\mu_B = 0$ MeV, while the bottom row [subplots (c) and (d)] is for a system with $\mu_B=400$ MeV, $\mu_I = -30$ MeV, $\mu_S=125$ MeV. The results at $\mu_B=0$ are compared with lattice QCD data from the HotQCD \cite{hot2} and WB collaboration \cite{borsanyi}, as well as with results from CMS ultra-central collisions \cite{cms} for subplot (a).}
	\end{figure}
	
	Further insight into the thermodynamic response of the medium can be gained from the temperature dependence of the speed of sound squared $c_s^2$ and the scaled specific heat $c_v/T^3$, which are shown in Fig. \ref{sound}. At $\mu_B=0$ (top row), the speed of sound exhibits a characteristic dip near the transition temperature. The location and depth of this minimum are sensitive to anisotropy. A positive $\xi$ makes the dip deeper and shifts it towards higher temperatures. In contrast, a negative $\xi$ shifts the dip to lower temperatures, consistent with the observed change in the chiral transition. The specific heat displays a corresponding peak in the transition region, which shifts to lower magnitude for positive $\xi$ and grows for negative $\xi$. Our results for both quantities at zero chemical potential are in qualitative agreement with available lattice data \cite{hot2,borsanyi}. At $\mu_B=400$ MeV (bottom row), these features are shifted to lower temperatures, with the softening of the equation of state being particularly evident.
	
	\begin{figure}
		\centering
		\begin{minipage}[c]{0.98\textwidth}
			(a)\includegraphics[width=7.4cm]{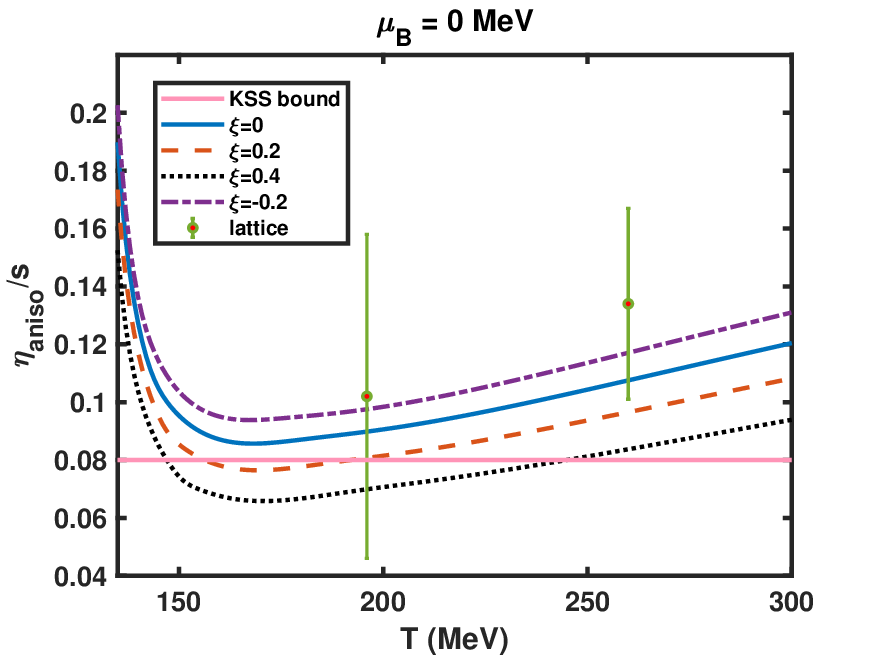}
			\hspace{0.03cm}
			(b)\includegraphics[width=7.4cm]{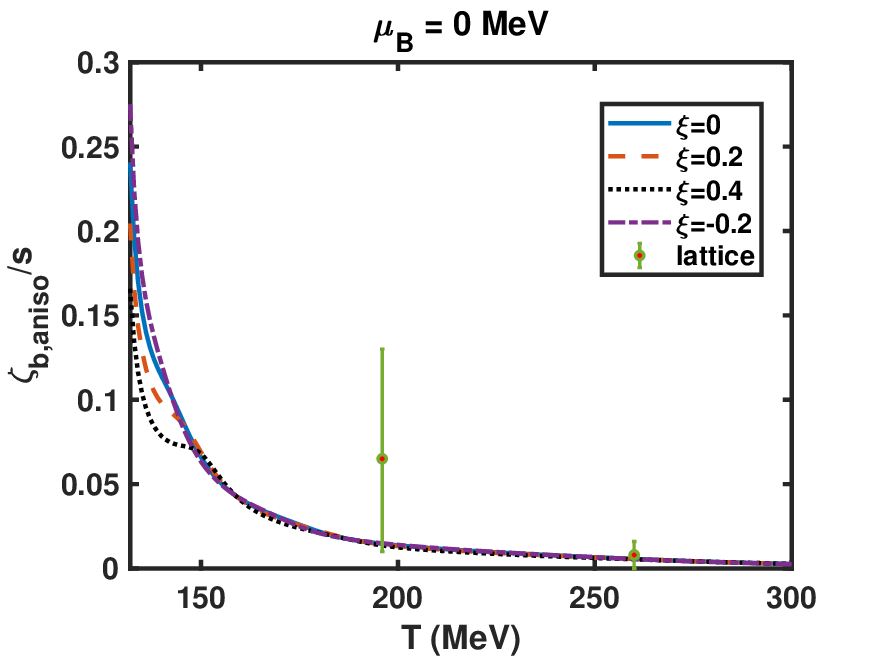}
			\hspace{0.03cm}	
			(c)\includegraphics[width=7.4cm]{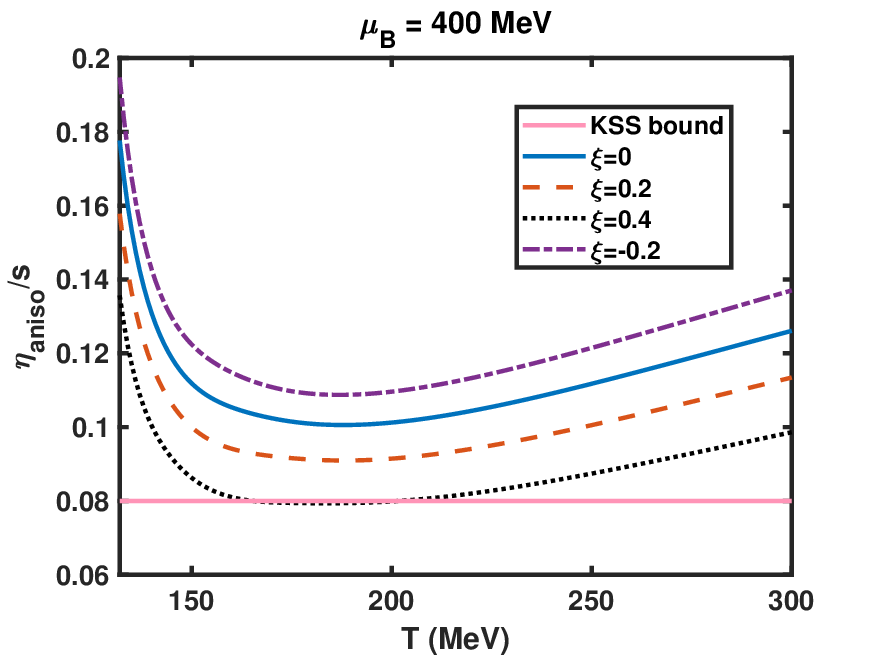}
			\hspace{0.03cm}
			(d)\includegraphics[width=7.4cm]{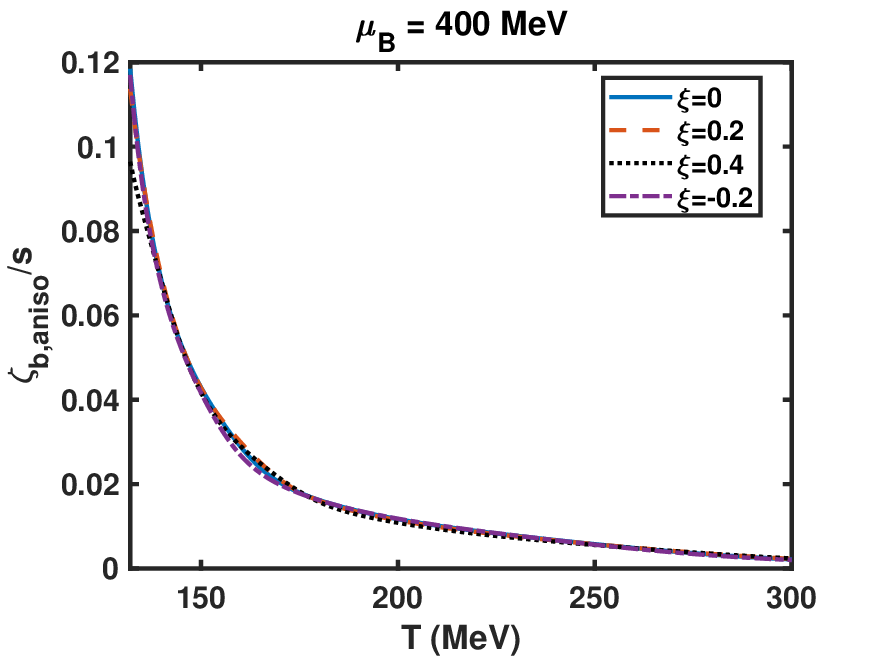}
			\hspace{0.03cm}	
		\end{minipage}
		\caption{\label{visc} The temperature dependence of specific shear viscosity $\eta_{\text{aniso}}/s$ and normalized bulk viscosity $\zeta_{b,\text{aniso}}$ for various values of anisotropy parameter $\xi=0$, 0.2, 0.4, and $-0.2$. The top row [subplots (a) and (b)] corresponds to a system at $\mu_B = 0$ MeV, while the bottom row [subplots (c) and (d)] is for a system with $\mu_B=400$ MeV, $\mu_I = -30$ MeV, $\mu_S=125$ MeV.}
	\end{figure}
	
	We now turn to the transport properties of the anisotropic quark matter. Fig. \ref{visc} presents the specific shear viscosity, $\eta_{\text{aniso}}/s$, and the normalized bulk viscosity, $\zeta_{b,\text{aniso}}/s$ as functions of temperature, for various $\xi$. Shear viscosity characterizes a fluid's resistance to shear stress, essentially its internal friction. A low shear viscosity allows the system to flow collectively with minimal energy loss, a property often described as ``perfect fluidity" \cite{romat07,lacey}. The value of $\eta$, especially when normalized by entropy density $s$ as the specific shear viscosity $\eta/s$, provides a direct measure of the interaction strength within the QGP. The specific shear viscosity, shown in Figs. \ref{visc}(a) for $\mu_B=0$ and \ref{visc}(c) for $\mu_B=400$ MeV, exhibits a minimum near the transition temperature, approaching the Kovtun-Son-Starinets (KSS) bound of $1/4\pi$ \cite{kss}, which is a hallmark of strongly coupled, nearly-perfect fluid. Below the transition, its value is large and decreases as temperature rises, while above the transition, it increases slowly with temperature. This is consistent with the finding across a wide range of theoretical frameworks, including the quasiparticle approach and various effective models \cite{sasaki,ghosh25,lata17,abhi,ghosh15,deb}. Momentum anisotropy has a significant impact; a positive $\xi$ (squeezed) decreases $\eta_{\text{aniso}}/s$, making the fluid less viscous and pushing the system even closer to the perfect fluid. This behavior is attributed to the growth of chromo-Weibel instabilities in the anisotropically expanding plasma. These instabilities generate turbulent color fields that provide a highly efficient, non-collisional mechanism for momentum transport, thereby reducing the effective shear viscosity \cite{lata17,chandra}. In contrast, a negative $\xi$ (stretched) increases $\eta_{\text{aniso}}/s$, resulting in a more viscous fluid. Such a prolate momentum configuration is expected to arise in the presence of a strong magnetic field, which constrains the motion of charged quarks to the lowest Landau level along the field direction \cite{rath19}. This is similar to the results obtained in the quasiparticle model approaches \cite{lata17,srivastava}, 2+1 flavor QM model \cite{zhang21}, and two flavor NJL model \cite{zhang22}. At a finite baryon chemical potential of $\mu_B = 400$ MeV (Fig. \ref{visc} (c)), the overall magnitude of $\eta_{\text{aniso}}/s$ increases across the temperature range, indicating that the system becomes more viscous at higher densities. Similar observations were reported in Ref. \cite{lata17}. The normalized bulk viscosity at $\mu_B=0$ MeV is plotted in Fig. \ref{visc} (b). This transport coefficient, which is a measure of the system's deviation from conformal invariance \cite{kharzeev,karsch}, is significant near the transition and diminishes at high temperatures, signaling the system's approach to the conformal limit where bulk viscosity is expected to vanish \cite{arnold06}. Our results for $\mu_B=0$ MeV (top row), which show a distinct peak in $\zeta_{b,\text{aniso}}/s$ in the transition region, are in good agreement with lattice calculations \cite{meyer2007,meyer2008}. The introduction of anisotropy has a minimal effect on $\zeta_{b,\text{aniso}}/s$ beyond the transition temperature. A squeezed momentum distribution $\xi>0$ slightly suppresses it, while a stretched distribution $\xi<0$ enhances it near the transition temperature. At higher temperatures, the effect of $\xi$ on $\zeta_{b,\text{aniso}}/s$ becomes negligible. As for the case of finite baryon chemical potential of $\mu_B=400$ MeV in Fig. \ref{visc} (d), we find that the magnitude of $\zeta_{b,\text{aniso}}/s$ is substantially suppressed compared to the $\mu_B=0$ case, and the effects of anisotropy become even less significant. This reduction at finite density is consistent with the general expectation that the system moves closer to conformality at higher densities for a given temperature.
	
	\begin{figure}
		\centering
		\begin{minipage}[c]{0.98\textwidth}
			(a)\includegraphics[width=7.4cm]{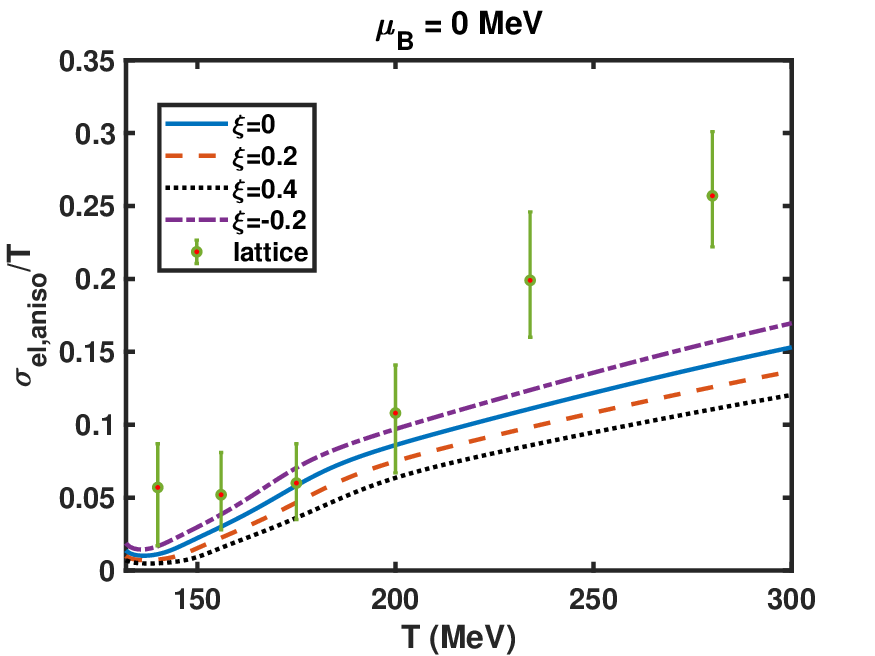}
			\hspace{0.03cm}
			(b)\includegraphics[width=7.4cm]{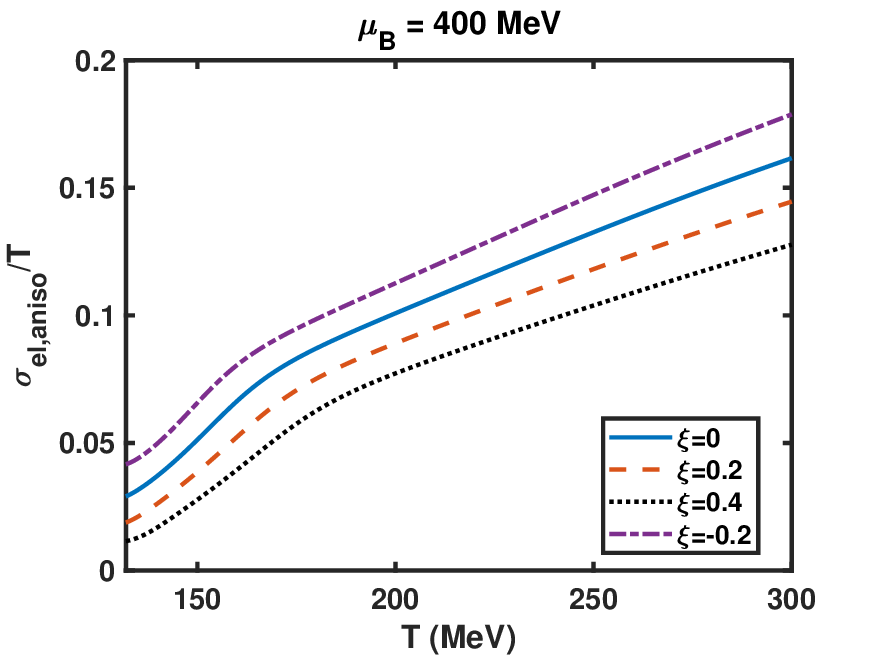}
			\hspace{0.03cm}
		\end{minipage}
		\caption{\label{elec} The temperature dependence of normalized electrical conductivity $\sigma_{el,\text{aniso}}/s$ for various values of anisotropy parameter $\xi=0$, 0.2, 0.4, and $-0.2$ at $\mu_B = 0$ MeV (a) and $\mu_B=400$ MeV, $\mu_I = -30$ MeV, $\mu_S=125$ MeV (b).}
	\end{figure}
	
	The effect of anisotropy on charge transport is examined via the normalized electrical conductivity, $\sigma_{el,\text{aniso}}/T$, which is shown as a function of temperature in Fig. \ref{elec}. The electrical conductivity increases with temperature as charge-carrying quarks become deconfined and more mobile. Our results for $\mu_B=0$ MeV (left panel) are in reasonable agreement with lattice data \cite{aarts}. The effect of anisotropy on conductivity is similar to its impact on shear and bulk viscosity. A positive $\xi$ (squeezed distribution), is found to suppress the electrical conductivity. This suppression becomes more significant as the anisotropy increases. This is consistent with the findings of Ref. \cite{rath19}, where the authors argue that an oblate momentum distribution hinders the transport of charge along the direction of the applied electric field. In contrast, a negative $\xi$ (stretched distribution) enhances the conductivity. This is similar to the findings in the 2+1 flavor QM model \cite{zhang21}, however contrasts the results in two flavor NJL model, where $\sigma_{el}/T$ first increases as $\xi$ increases at low temperatures, but as temperature rises further, the values of $\sigma_{el}/T$ for different $\xi$ eventually overlap \cite{zhang22}. At finite baryon chemical potential (right panel), the overall magnitude of the electrical conductivity is higher than at $\mu_B=0$. This is a direct consequence of the increased net charge carrier density (an excess of quarks over antiquarks) at finite $\mu_B$. Importantly, the relative effect of $\xi$ remains the same as in the vanishing chemical potential case. Similar findings were reported in quasiparticle model approaches in Refs. \cite{srivastava} and \cite{lata17}.
	
	\begin{figure}
		\centering
		\begin{minipage}[c]{0.98\textwidth}
			(a)\includegraphics[width=7.4cm]{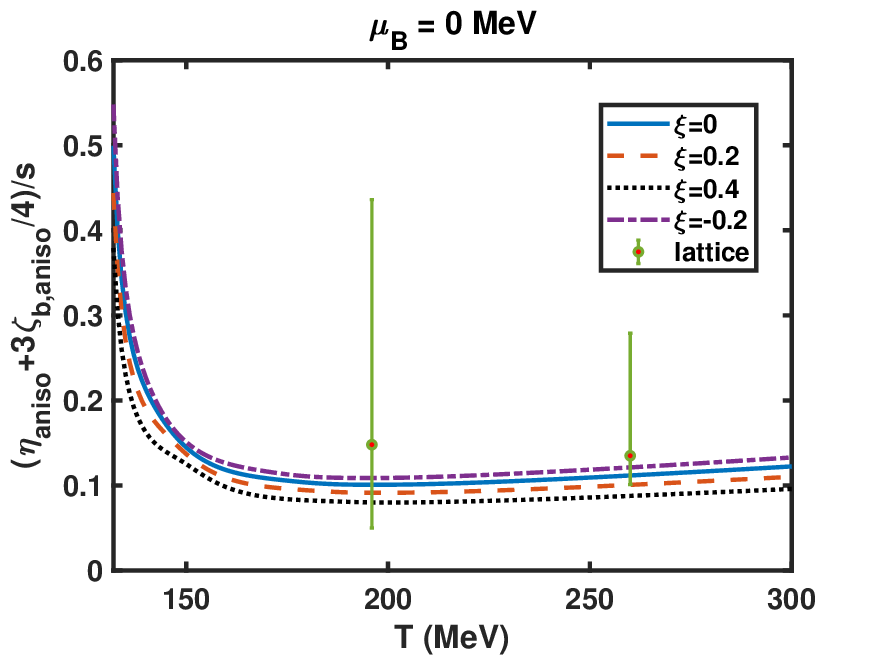}
			\hspace{0.03cm}
			(b)\includegraphics[width=7.4cm]{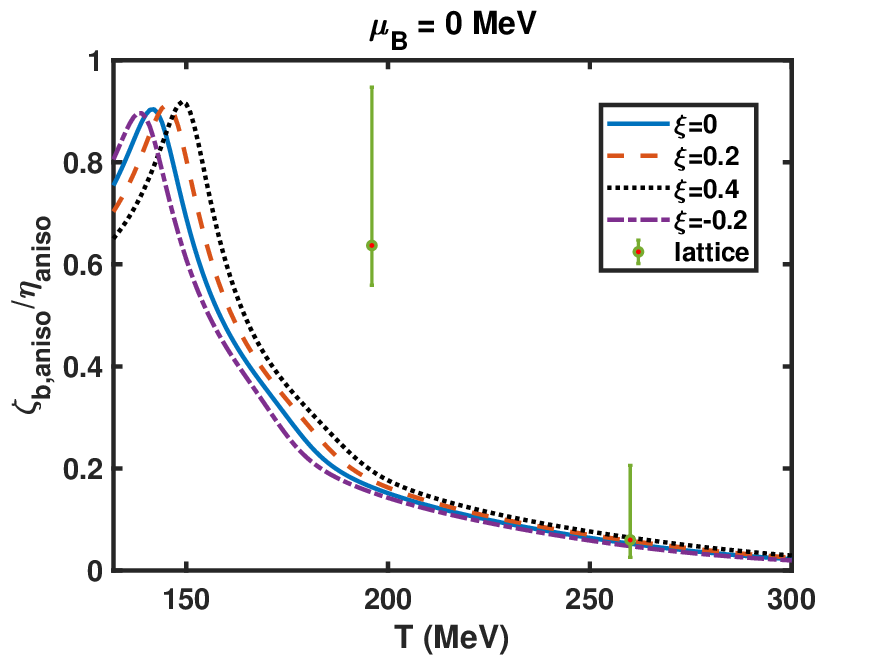}
			\hspace{0.03cm}	
			(c)\includegraphics[width=7.4cm]{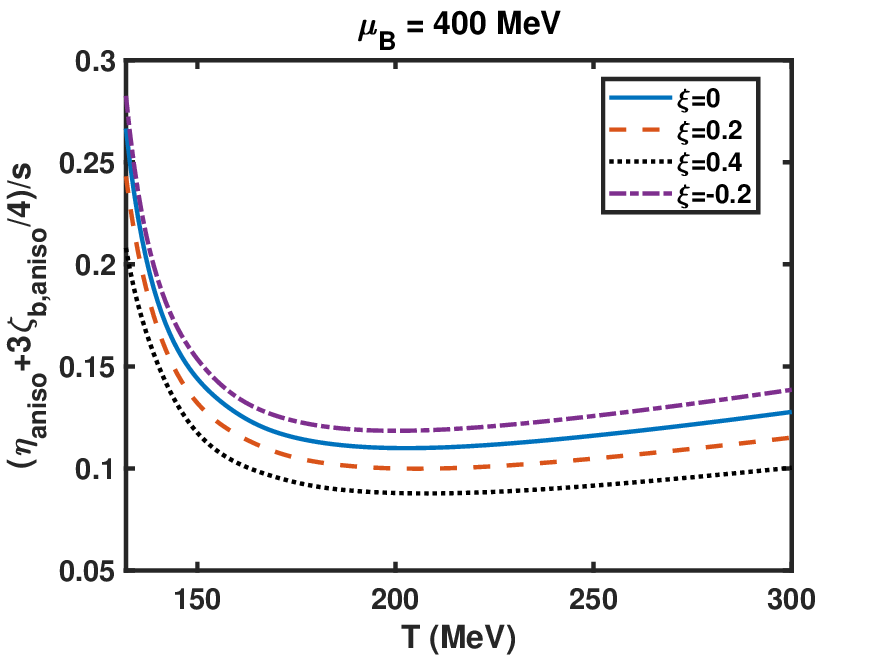}
			\hspace{0.03cm}
			(d)\includegraphics[width=7.4cm]{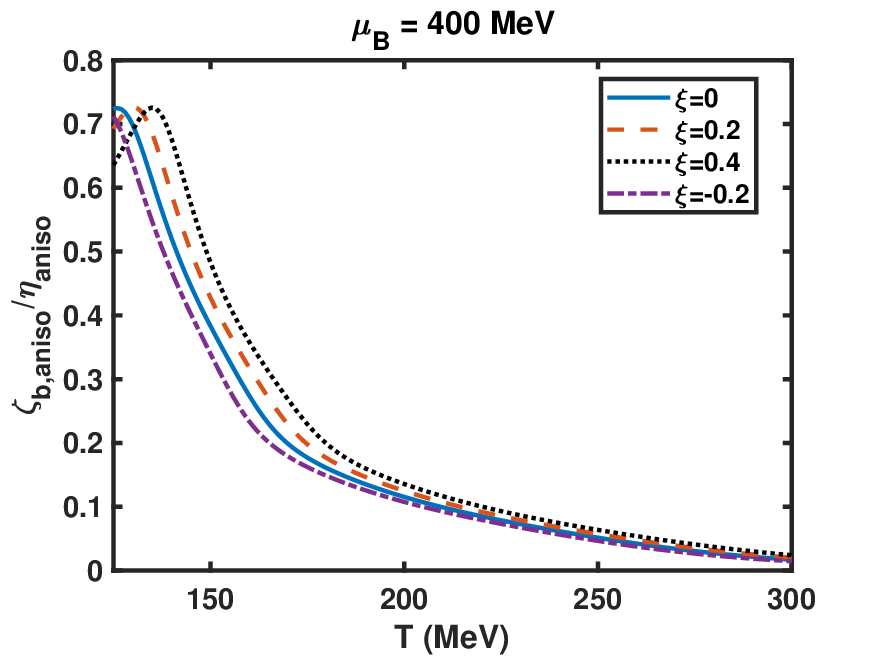}
			\hspace{0.03cm}	
		\end{minipage}
		\caption{\label{ratio} The variation of specific sound channel $(\eta+3\zeta_b/4)/s$ and bulk to shear viscosity ratio $\zeta_b/\eta$ with temperature for various values of anisotropy parameter $\xi=0$, 0.2, 0.4, and $-0.2$. The top row corresponds to a system at $\mu_B = 0$ MeV, while the bottom row is for a system with $\mu_B=400$ MeV, $\mu_I = -30$ MeV, $\mu_S=125$ MeV.} 
	\end{figure}
	
	Finally, in Fig. \ref{ratio}, we display  the thermal dependence of two important quantities that characterize the dissipative nature of the medium: the specific sound channel, $(\eta_{\text{aniso}}+3\zeta_{b,\text{aniso}}/4)/s$, and the ratio of bulk to shear viscosity, $\zeta_{b,\text{aniso}}/\eta_{\text{aniso}}$, for various $\xi$ values. The behavior of the specific sound channel (left panel) is largely governed by the specific shear viscosity, as the bulk viscosity contribution is negligible in comparison across most of the temperature range, except in the vicinity of the transition region, as shown in Fig. \ref{visc} (b) and (d). Consequently, the specific sound channel exhibits a minimum near $T_c$, and its dependence on $\xi$ directly mirrors that of $\eta_{\text{aniso}}/s$. The bulk-to-shear ratio, $\zeta_{b,\text{aniso}}/\eta_{aniso}$, provides a measure of the interplay between bulk and shear viscous effects. This ratio exhibits a prominent peak in the transition region at $\mu_B=0$ MeV (Fig. \ref{ratio}(b)), indicating that bulk viscosity is most significant during the phase transition. Momentum anisotropy influences the location of this peak: positive anisotropy ($\xi > 0$) shifts this peak to higher temperatures, while negative anisotropy ($\xi < 0$) shifts it to lower $T$. At $\mu_B=400$ MeV (Fig. \ref{ratio}(d)), the peak becomes less pronounced and shifts to lower temperatures. This suppression suggests a faster approach to conformal symmetry as baryon density increases. This finding is consistent with the general expectation that the relative importance of bulk viscosity diminishes as the system moves away from the crossover region into the denser, more weakly interacting regime.
	

	\section{Summary}
	\label{summary}
	In this work, we have presented a comprehensive analysis of the impact of momentum-space anisotropy on the thermodynamic equation of state, and transport properties of strongly interacting quark matter. Employing the Polyakov chiral SU(3) quark mean field (PCQMF) model, which captures both chiral symmetry dynamics and quark confinement through a Polyakov loop potential, we introduce anisotropy via a spheroidal deformation in the momentum distribution, quantified by the parameter $\xi$. A system with a ``squeezed" or oblate momentum distribution ($\xi>$0) preferentially populates transverse momentum states, characteristic of the rapid longitudinal expansion in the early stages of heavy-ion collisions. While a ``stretched" or prolate distribution ($\xi<$0) favors momenta along the anisotropy axis, a scenario expected to arise in the presence of a strong magnetic field, with $\xi=$0 representing the isotropic case. Our investigation reveals that even weak anisotropy profoundly alters the system's characteristics. 
	We find that key thermodynamic observables like pressure, energy density, entropy density, squared speed of sound, and specific heat are systematically suppressed for squeezed systems ($\xi>$0) and enhanced for stretched ones ($\xi<$0) when compared to the isotropic case at the same temperature. This demonstrates that the momentum distribution fundamentally modifies the available phase space and thus alters the bulk thermodynamic properties and the equation of state of the medium.
	
	Furthermore, the transport coefficients, calculated by solving the relativistic Boltzmann equation in the relaxation time approximation, exhibit a strong and systematic dependence on this anisotropy. The specific shear viscosity ($\eta_{\text{aniso}}/s$) and electrical conductivity ($\sigma_{el,\text{aniso}}/T$) are significantly suppressed for squeezed distributions ($\xi>$0), pushing the system closer to the ``perfect fluid" ideal and hindering charge transport, respectively. Conversely, these coefficients are enhanced for stretched distributions ($\xi<$0). The bulk viscosity ($\zeta_{b,\text{aniso}}/s$) shows a similar, though less pronounced, trend, with its characteristic peak near the phase transition shifting in temperature and magnitude according to the value of $\xi$. At a finite baryon chemical potential of $\mu_B=400$ MeV, the absolute magnitudes of $\eta_{\text{aniso}}/s$ and $\zeta_{b,\text{aniso}}/s$ change, with shear viscosity increasing while bulk viscosity is suppressed. However, the qualitative dependence on the sign of $\xi$ remains the same. Collectively, these results underscore that momentum-space anisotropy is a crucial physical feature that fundamentally reshapes the properties of the quark-gluon plasma. Even when weak, it induces significant modifications in both thermodynamic and transport properties of the QCD medium. The thermodynamic responses are suppressed for positive $\xi$. The transport coefficients are modified significantly with $\xi$, especially near the transition temperature, highlighting the anisotropic redistribution of microscopic transport. Its inclusion is therefore essential for the realistic phenomenological modeling of the early-time, pre-equilibrium dynamics in ultrarelativistic heavy-ion collisions at facilities like RHIC, the LHC, FAIR, and NICA, and for accurately bridging theoretical predictions with experimental data.
	
	\section{ACKNOWLEDGMENT}
	DS sincerely acknowledges the support for this work from the Ministry of Science and Human Resources (MHRD), Government of India, through an Institute fellowship under the National Institute of Technology Jalandhar. AK sincerely acknowledges Anusandhan National Research Foundation (ANRF), Government of India, for funding the research project under the Science and Engineering Research Board-Core Research Grant (SERB-CRG) scheme (File No. CRG/2023/000557).
	
	\bibliographystyle{apsrev4-2}

\end{document}